\documentclass{article}%
\usepackage{amsmath}
\usepackage{amsfonts}
\usepackage{amssymb}
\usepackage{graphicx}%
\begin{document}
{\huge Role of general relativity }

{\huge and quantum mechanics }

{\huge in dynamics of Solar System }

$\ \ \ \ \ \ \ \ \ \ \ \ \ \ \ \ \ \ \ \ \ \ \ \ \ \ \ \ \ \ \ \ \ $

\ \textbf{Arkady L. Kholodenko}

375 H.L.Hunter Laboratories,

Clemson\textit{\ }University, Clemson, SC 29634-0973, USA

E-mail: string@clemson.edu

\medskip\ \ \ \ \ \ \ \ \ \ \ \ \ \ \ \ \ \ \ \ \ \ \ \ \ \ \ \ \ \ \ \ \ \ \ \ \ \ \ \ \ \ \ \ \ \ \ \ \ \ \ \ \ \ \ \ \ \ \ \ \ \ \ \ \ \ \ \ \ \ \ \ \ \ \ \ \ \ \ \ \ \ \ \ \ \ \ \ \ 

\textbf{Abstract}

Let $m_{i}$ be the mass of $i$-th planet and $M_{\odot}$ be the Solar mass. From

astronomical data it is known that ratios $r_{i}=m_{i}/(m_{i}+M_{\odot})$ are of

order $10^{-6}-10^{-3}$ for all planets. The same is true for almost all

satellites of heavy planets, with exception of Moon-Earth for which it is
$10^{-2}$

and Charon-Pluto, for which it is $10^{-1}.$ These  results strongly suggest 

that Einstein's treatment of the Mercury dynamics can be safely

extended to almost any object in the Solar System. If this is done,

gravitational interactions between planets/satellites can be ignored since

they move on geodesics. This fact \ still does not explain the existing order

in the Solar System. Because of it, all planets lie in the same (Suns's

equatorial) plane and move in the same direction coinciding with that for

the rotating Sun. The same is true for the regular satellites of heavy planets 

and for the planetary ring systems associated with these satellites.

Such filling pattern is typically explained with help of the hypothesis by

which our Solar System is a product of evolutionary dynamics of some

pancake-like cloud of dust. This hypothesis would make sense should the

order in our planetary system (and that for exoplanets \ rotating around

other stars) be exhausted by the pattern just described. But it is not!

In addition to regular satellites there \ are irregular satellites (and at least

one irregular (Saturn) ring associated with such \ a satellite (Phoebe)) grouped

in respective planes (other than equatorial) in which they all move in

wrong directions on stable orbits. These are located strictly outside of those

for regular satellites. Since this filling pattern is reminiscent of filling

patterns in atomic mechanics, based on the original Heisenberg's ideas,

we develop quantum \ celestial mechanics explaining this pattern. In such a

mechanics the Plank's constant is replaced by another constant, different

for each planetary/satellite system. To test correctness of \ our theory the

number of allowed stable orbits for planets and for regular satellites of

heavy planets is calculated resulting in good \ agreement with \ observational

data. Developed formalism takes essentially into account the fact that 

all planets and \ most all satellites are moving on geodesics. In addition, the

paper briefly discusses quantum mechanical nature of \ rings of heavy planets

and potential usefulness of the obtained results for cosmology.\ \ \ \ \ \ \ \ 


\section{Introduction}

\subsection{Background}

The impact of quantum mechanics on physics requires no comments. More
surprising is the impact of quantum mechanical thinking on mathematics and
mathematicians. According to world renown mathematician, Yuri Manin [1], the
traditional view "continuous from discrete" gives (now) a way to the inverted
paradigm: "discrete from continuous". This inverted paradigm in mathematics
brought to life quantum logic, deformation quantization, quantum topology,
quantum cohomology, etc. Quantum q-calculus beautifully described in the book
by Kac and Cheung [2] provide a nice example of unified view of continuous and
discrete \ using examples from analysis and theory of orthogonal polynomials
(used essentially in quantum mechanics). These results are sufficient for, at
least formal, looking at results of classical celestial mechanics from such
q-deformed (quantum) point of view. Surprisingly, the "quantization process"
had already began in celestial mechanics. Usefulness of quantum mechanics has
been recently discussed in [3-5]. This paper is aimed at extension of the
emerging trend. We believe, that there are reasons much deeper than just a
current fashion for such an extension.

In his "Les Methodes Nouvelles de la Mecanique Celeste" written \ between 1892
and 1899 [6] Poincare$^{\prime}$ developed theory of celestial mechanics by
assuming that all planets, and \ all satellites of heavy planets are rotating
in the same direction coinciding with direction of rotation of the Sun around
its axis. To a large degree of accuracy all planets lie in the same (Sun's
equatorial) plane. These assumptions were legitimate in view of the
astronomical data available to Poincare$^{\prime}$. These data suggest that
known to us now Solar System most likely is result of dynamic evolution of the
pancake-like rotating self-gravitating cloud of dust [7]. In 1898 the shocking
\ counter example to the Poincare$^{\prime}$ theory was announced by Pickering
who discovered the ninth moon of Saturn (eventually named Phoebe) rotating in
the direction opposite to all other satellites of Saturn. \ Since that time
the satellites rotating in the "normal" direction are called "\textit{regular}%
" (or "\textit{prograde}") while those rotating in the opposite direction
called "\textit{irregular}"(or "\textit{retrograde}"). At the time of writing
of this paper 103 irregular satellites were discovered (out of those, 93 were
discovered after 1997 thanks to space exploration by rockets)\footnote{E.g.
read "Irregular moon" in Wikipedia}. Furthermore, in the late 2009 \ Phoebe
had brought yet another surprise to astronomers. Two articles in Nature [8,9]
are describing the largest new ring of Saturn. This new ring lies in the same
plane as Phoebe's orbit and, in fact, the \ Phoebe's trajectory is located
inside of the ring. The same arrangement is true for regular satellites and
the associated with them rings.

From the point of view of Newton's laws of gravity there is no reason to
restrict trajectories of planets to the same plane or to expect that all
planets will rotate in the same direction. Their observed locations thus far
are attributed to the conditions at which the Solar System was born. This
assumption is plausible and is in accord with the model \ (and its
interpretation) of Solar System used by Poincare$^{\prime}.$ But, as stated
above, the major exception occurred already in 1898 when Poincare$^{\prime}$
was about to finish his treatise. To repair the existing theory one has to
make an assumption that all irregular satellites are "strangers". That is that
they were captured by the already existing \ and fully developed Solar System.
Such an explanation \ would make perfect sense should the orbits of these
strangers be arranged in a completely arbitrary fashion. But they are not!
Without an exception it is known that: a) all retrograde satellite orbits are
lying \textsl{strictly outside} of the orbits of prograde satellites, b) the
inclinations of their orbits is noticeably different from \ those for prograde
satellites, however, c) by analogy with prograde satellites they tend to group
(with few exceptions) in orbits-all having the same inclination so that
different groups of retrogrades are having differently inclined orbits in such
a way that these orbits \ do not overlap if the retrograde plane of
\ satellites with one inclination is superimposed with that for another
inclination\footnote{E.g. see http://nssdc.gsfc.nasa.gov/planetary/ then, go
to the respective planet and, then-to the "fact sheet" link for this planet.}.
In addition, all objects lying outside the sphere made by the rotating plane
in which all planets lie are arranged in a similar fashion[10]. Furthermore,
the orbits of prograde satellites of all heavy planets \ lie in the respective
equatorial planes- just like the Sun and the planets - thus forming miniature
Solar-like systems. These equatorial planes are tilted with respect to the
Solar equatorial plane since all axes of rotation of heavy planets are
tilted\footnote{That is the respective axes of rotation of heavy planets are
not perpendicular to the Solar equatorial plane.} with different angles for
different Solar-like systems. These "orderly" facts make nebular origin of our
Solar System questionable. To strengthen the doubt further we would like to
mention that for the exoplanets\footnote{E.g. see http://exoplanets.org/} it
is not uncommon to observe planets rotating in the "wrong" direction around
the respective stars\footnote{E.g. read \ "Retrograde motion" in Wikipedia}.
This trend goes even further to objects such as galaxies. In spiral galaxies
the central bulge typically co-rotates with the disc. But for the galaxy
NGC7331 the bulge and the disc are rotating in the opposite directions. These
facts bring us to the following subsection.

\subsection{Statements of problems to be solved}

From the discussion above it looks like there is some pattern of filling of
the orbits in Solar and Solar-like systems. First, the prograde orbits are
being filled-all in the same equatorial plane. Second, the retrograde orbits
start to fill in-also in respective planes tilted with respect to equatorial
(prograde) plane. These tilted planes can be orderly arranged by the observed
typical distance for retrogrades satellites: those lying in different planes
\ will have different typical distances to the planet around which they
rotate. All these retrograde orbits (without an exception!) are more distant
from the respective planets than the prograde orbits. Inclusion of rings into
this picture does not change the pattern just described. All heavy planets do
have system of rings. These are located in the respective equatorial planes.
The newly discovered Phoebe ring fits this pattern perfectly. While the rings
associated with the prograde satellites all live in the respective equatorial
planes which are "correctly" rotating , the \ Phoebe ring is rotating in
"wrong" direction and is tilted (along with Phoebe's trajectory) by 27$^{0}$
with respect to Saturn's equatorial plane [8,9]. Given all these facts, we
would like to pose the following

\textbf{Main Question}: \textsl{What all these just noticed filling patterns
have to do with general relativity}? \ \ \ \ \ \ \ \ \ \ \ \ 

In this paper we would like to argue that, in fact, to a large extent the
observed patterns are manifestations of effects of general relativity at the
length scales of our Solar System. Indeed, let $M_{\odot}$ be the mass of the
Sun (or, respectively, heavy planet such as Jupiter, Saturn, etc.) and $m_{i}$
be the mass of an i-th planet(respectively, the i-th satellite of heavy
planet). Make a ratio $r_{i}=\frac{m_{i}}{m_{i}+M_{\odot}}$ . The analogous
ratios can be constructed for respective heavy planets (Jupiter, Saturn,
Uranus, Neptune) and for any of their satellites. The observational data
indicate that with only two exceptions: Earth-Moon (for which the ratio
$r\sim10^{-2}),$ and Pluto-Charon (for which the ratio $r\sim10^{-1}),$ all
other ratios in the Solar System are of order $10^{-6}-10^{-3}$[11\textbf{]}.
Everybody familiar with classical mechanics knows that under such
circumstances the center of mass of such a binary system practically coincides
with that for $M_{\odot}$. And if this is so, then the respective trajectories
can be treated as geodesics. Hence, not only motion of the Mercury can be
treated in\ this way, as it was done by Einstein, but also motion of almost
any satellite\footnote{Regrettably, not our Moon! \ Description of dynamics of
Moon is similar to that for rings of heavy planets (to be discussed in Section
4) and, as such, is also quantizable.} in the Solar System! \ Such a
replacement of Newton's mechanics by Einsteinian mechanics of general
relativity even though plausible but is still not providing us with the answer
to the Main Question. Evidently, the observed mass ratios and the observed
filling patterns must have something in common. If we accept the point of view
that the observed filling patterns are possible if and only if the observed
mass ratios allow us to use the Einstein's geodesics, then we inevitably
arrive at quantization of Solar System dynamics. \ Such a statement looks
rather bizarre since the traditional quantum mechanics is dealing with
microscopic objects. Nevertheless, as results of Refs.[3-5] indicate, the
formalism of quantum mechanics can indeed be adopted to problems emerging in
celestial mechanics. Thus, we arrive at the statements of problems to be
studied in the main text.

First, we need to prove that Einsteinian relativity favors quantum mechanical
description of Solar System dynamics. Second, we need to prove that such
quantum mechanical description is capable of explaining the observed \ in
Solar System filling patterns. Evidently, the combined solutions of the first
and second problems provide an answer to the Main Question.

\subsection{Organization of the rest of the paper}

This paper contains 5 sections and 3 appendices. In Section 2 we discuss
historical, mathematical and physical reasons for quantization of the Solar
System. In particular, by using some excerpts from works by Laplace and
Poincare$^{\prime}$ we demonstrate that Laplace can be rightfully considered
as founding father of both general relativity and quantum mechanics. He used
basics of both of these disciplines in his study of dynamics of \ known at
that time satellites of Jupiter. Specifically, in his calculations masses of
satellites were ignored and when they were made nonzero but small the
(Einsteinian) orbits were replaced by those which form standing waves around
the Einsteinian orbits. Mathematical comments made by Poincare$^{\prime}$ on
Laplace's work are essentially same as were later unknowingly used by
Heisenberg in his formulation of quantum mechanics. \ In Section 3 we discuss
some changes in the existing apparatus of quantum mechanics needed for
development of physically meaningful quantization of Solar System dynamics.
The SO(2,1) symmetry typical for planar configurations is investigated in
detail so that amendment to traditional quatization scheme remain compatible
with this symmetry group.\ In Section 4 we use this amended formalism for
description of Solar System dynamics and explanation of the empirically
observed filling patterns. The main results are summarized in Table 2 \ and in
Subsection 4.2. In Table 2 we compare our calculation of available orbits for
planets and for regular satellites of heavy planets with empirically observed.
Obtained theoretical results are in reasonable accord with empirically
observed and with the quantum mechanical rules for the filling of orbits
discussed in Subsection 4.2. These results are extended in Subsection 4.3
describing rationale for quantization of dynamics of rings around heavy
planets. In Section 5 we discuss the problem of embedding of the Lorentzian
group SO(2,1) into larger groups such as SO(3,1), SO(4,1), SO(4,2), etc. This
is done with assumption that these larger symmetries should be taken into
account in anticipation that quantum dynamics of Solar System could be
eventually used in testing some cosmological models/theories. Paper concludes
with Section 6 in which we discuss \ the reasons why the developed formalism
fits the combinatorial theory of group representations recently discussed by
mathematicians Knutson and Tao[12-14]\footnote{See also
www.math.ucla.edu/\symbol{126}tao/java/Honeycomb.html} and applied to quantum
mechanical problems by Kholodenko [15,16]. Appendices A-C supplement some
results of Sections 3 and 4.

\section{Harmonious coexistence of general relativity and quantum mechanics in
the Solar System}

\subsection{From Laplace to Einstein and via Poincare$^{^{\prime}}$}

Everybody knows that Einstein considered the Copenhagen version of quantum
mechanics as incomplete/temporary. He was hoping for a deeper \ quantum theory
in which God is not playing dice. His objections, in part, had been caused by
the fact that the "new" quantum mechanics and the "old" general relativity
have nothing in common. In this section we would like to argue, that such an
attitude by Einstein is caused, most likely, by circumstances of his life and
that \textsl{these two disciplines actually have the same historical origin}.
It is known that Einstein was not too excited about the works of Henry
Poincare$^{^{\prime}}$and, in return, Poincare$^{^{\prime}}$ never quoted
works by Einstein\footnote{Only late in his life Einstein did acknowledged
Poinancare's contributions to science [17]}. As result of this historical
peculiarity, in his seminal works on general relativity Einstein never quoted
Poincare's revolutionary results on celestial mechanics. Thus, the celebrated
shift of Mercury's\ perihelium was obtained entirely independent of Poincare's
results! Correctness of Einsteinian relativity had been tested many times,
including results obtained in 2008 and 2010 [18,19]. These latest results are
the most accurate to date. They unambiguously support general relativity in
its canonical form at least at the scales of our Solar System. \ Einstein's
victory over Poincare$^{^{\prime}}$ is mysterious in view the following
\ facts from celestial mechanics. To discuss these facts, we need to provide
some background from classical mechanics first. In particular, even though
classical Hamiltonians for Coulombic and Newtonian potentials look
\textsl{almost} the same, they are far from being \textsl{exactly} the same.
In the classical Hamiltonians for multielectron atoms \ all electron masses
are the same, while for the Solar-like planetary system the masses of all
satellites are different. \ In some instances to be discussed below such
difference can be made non existent. In such cases formal quantization for
both systems \ can proceed in the same way. To explain how this happens, we
begin with two- body Kepler problem treated in representative physics
textbooks [20]. Such treatments tend to ignore the equivalence principle-
essential for the gravitational Kepler problem and nonexistent for the
Coulomb-type problems. Specifically, the description of general relativity in
Vol.2 of the \ world-famous Landau-Lifshitz course in theoretical physics [21]
begins with the Lagrangian for the particle in gravitational field $\varphi$:
$\mathcal{L}$=$\dfrac{m\mathbf{v}^{2}}{2}-m\varphi.$ The Newton's equation for
such a Lagrangian reads:
\begin{equation}
\mathbf{\dot{v}=-\nabla}\varphi\mathbf{.} \tag{2.1}%
\end{equation}
Since the mass drops out of this equation, it is possible to think about such
an equation as an equation for a geodesic in (pseudo)Riemannian space. \ This
observation, indeed, had lead Einstein to full development of theory of
general relativity and to his calculation of the Mercury's perihelion shift.
The above example is misleading though. Indeed, let us discuss the 2-body
Kepler problem for particles with masses $m_{1}$ and $m_{2}$ interacting
gravitationally. The Lagrangian for this problem is given by%
\begin{equation}
\mathcal{L}=\frac{m_{1}}{2}\mathbf{\dot{r}}_{1}^{2}+\frac{m_{2}}%
{2}\mathbf{\dot{r}}_{2}^{2}+\gamma\frac{m_{1}m_{2}}{\left\vert \mathbf{r}%
_{1}-\mathbf{r}_{2}\right\vert }. \tag{2.2}%
\end{equation}
Introducing, as usual, the center of mass \ and relative coordinates via
$m_{1}\mathbf{r}_{1}+m_{2}\mathbf{r}_{2}=0$ and $\mathbf{r}=\mathbf{r}%
_{1}-\mathbf{r}_{2},$ the above Lagrangian \ acquires the following form:%
\begin{equation}
\mathcal{L=}\frac{\mu}{2}\mathbf{\dot{r}}^{2}+\gamma\frac{m_{1}m_{2}%
}{\left\vert \mathbf{r}\right\vert }\equiv\frac{m_{1}m_{2}}{m_{1}+m_{2}}%
(\frac{\mathbf{\dot{r}}^{2}}{2}+\gamma\frac{(m_{1}+m_{2})}{\left\vert
\mathbf{r}\right\vert }), \tag{2.3}%
\end{equation}
where, as usual, we set $\mu=\frac{m_{1}m_{2}}{m_{1}+m_{2}}.$The constant
$\frac{m_{1}m_{2}}{m_{1}+m_{2}}$ can be dropped and, after that, instead of
the geodesic, Eq.(2.1), we obtain the equation for a fictitious point-like
object of unit mass moving in the field of gravity produced by the point-like
body of mass $m_{1}+m_{2}$. Clearly, in general, one cannot talk about
geodesics in this case even though Infeld and Schild had attempted to do just
this already in 1949 [22].\ The case is far from being closed even in 2010
[23]. \ These efforts look to us mainly as academic (unless dynamics of binary
stars is considered) for the following reasons. If, say, $m_{1}\gg m_{2\text{
}}$ as for the electron in Hydrogen atom or for the Mercury rotating around
Sun \ one can (to a very good accuracy) discard mass $m_{2\text{ }}$ thus
obtaining the equation for a geodesic coinciding with (2.1). In the
Introduction we defined the ratio $r=\frac{m_{2}}{m_{1}+m_{2}}.$ If we do not
consult reality for guidance, the ratio $r$ can have \textsl{any} nonnegative
value. However, what is observed in the sky (and in atomic systems as well)
leads us to the conclusion that (ignoring our Moon) all satellites of heavy
planets as well as all planets of our Solar System move along geodesics
described by (2.1), provided that we can ignore interaction between the
planets/satellites. We shall call such an approximation the \textsl{Einstein's
limit. }It is analogous to the mean field Hartree-type approximation in atomic
mechanics. If we believe Einstein, then such Hartree-type approximation does
not require any corrections. This looks like "too good to be true". Indeed,
the first who actually used Einstein's limit (more then 100 years before
Einstein!) in his calculations was Laplace [24]\textbf{,} Vol.4. \ In his book
[6], Vol.1, art 50, Poincare$^{^{\prime}}$ discusses Laplace's work on
dynamics of satellites of Jupiter. Incidentally, Laplace also studied motion
of the Moon, of satellites of Saturn and Uranus and of Saturn's ring system
[24]\textbf{,}Vol.s 2,4. \ 

Quoting from Poincare$^{^{\prime}}$:

"(Following Laplace) consider the central body of large mass \ (Jupiter) and
three other small bodies (satellites Io, Europe and Ganymede), \textsl{whose
masses} \textsl{can be taken to be zero, }rotating around a large body in
accordance with Kepler's law. Assume further that the eccentricities and
inclinations of the orbits of these (zero mass) bodies are equal to zero, so
that the motion is going to be circular. Assume further that the frequencies
of their rotation $\omega_{1},\omega_{2}$ and $\omega_{3\text{ }}$are such
that there is a linear relationship%
\begin{equation}
\alpha\omega_{1}+\beta\omega_{2}+\gamma\omega_{3\text{ }}=0 \tag{2.4}%
\end{equation}
with $\alpha,\beta$ and $\gamma$ being three mutually simple integers such
that
\begin{equation}
\alpha+\beta+\gamma=0. \tag{2.5}%
\end{equation}
Given this, it is possible to find another three integers $\lambda
,\lambda^{\prime}$ and $\lambda^{\prime\prime}$ such that $\alpha\lambda
+\beta\lambda^{\prime}+\gamma\lambda^{\prime\prime}=0$ implying that
$\omega_{1}=\lambda A+B,\omega_{2}=\lambda^{\prime}A+B,\omega_{3}%
=\lambda^{\prime\prime}A+B$ with $A$ and $B$ being some constants. After some
time $T$ \ it is useful to construct the angles \ $T(\lambda A+B),T(\lambda
^{\prime}A+B)$ and $T(\lambda^{\prime\prime}A+B)$ describing current location
of respective satellites (along their circular orbits) and, their differences:
$(\lambda-\lambda^{\prime})AT$ and $(\lambda-\lambda^{\prime\prime})AT.$ If
now we choose $T$ in such a way that $AT$ is proportional to $2\pi,$ then the
angles made by the radius-vectors (from central body to the location of the
planet) will coincide with those for $T=0$. Naturally, such a motion (with
zero satellite masses) is periodic with period $T$.

\textsl{The question remains: Will the motion remain periodic in the case if
masses are small but not exactly zero? That is, if one allows the satellites
to interact with each other?.}...

\textsl{Laplace had demonstrated that the orbits of these three satellites of
Jupiter will differ only slightly from truly periodic. In fact, the locations
of these satellites are oscillating around the zero mass trajectory}"

Translation of this last paragraph into language of modern quantum mechanics
reads: Laplace demonstrated that only the Einsteinian trajectories are subject
to the Bohr-Sommerfel'd- type quantization condition. \textbf{That is at the
scales of Solar System correctness of Einsteinian general} \textbf{relativity
is assured by correctness of quantum mechanics (}closure of the
Laplace-Lagrange oscillating orbits [25]\textbf{) and vice-versa so that these
two theories are inseparably linked together}.\ 

The attentive reader of this excerpt from Poincare$^{^{\prime}}$ could already
realized that Laplace came to such a conclusion based on (2.4) as starting
point. Thus, the condition (2.4) can be called quantization condition (since
eventually it leads to the Bohr-Sommerfel'd condition). \ Interestingly
enough, this condition was chosen by Heisenberg [26] as fundamental
quantization condition from which all \ machinery of quantum mechanics can be
deduced! This topic is further \ discussed in the next subsection. Before
doing so we notice that extension of work by Laplace to the full $n+1$ body
planar problem was made only in 20th century and can be found in the monograph
by Charlier [27]. More rigorous mathematical proofs involving KAM theory have
been obtained just recently by Fejoz [28] and Biasco et al [29]. The
difficulty, of course, is caused by proper accounting of the effects of finite
but nonzero masses of satellites and \ by showing that, when these masses are
very small, the Einsteinian limit makes perfect sense and is stable. \ A
sketch of these calculations for planar four-body problem ( incidentally
studied by de Sitter in 1909) can be found in nicely written lecture notes by
Moser and Zehnder [30].

\subsection{From Laplace to Heisenberg and beyond}

Very much like Einstein, who without reading of works by Poincare on celestial
mechanics arrived at correct result for dynamics of Mercury, Heisenberg had
arrived at correct formulation of quantum mechanics without reading works by
both Poincare$^{^{\prime}}$ and Laplace$^{^{\prime}}$. In retrospect, this is
not too surprising: correctly posed problems should lead to correct solutions.
In the case of Einstein, his earlier obtained result $E=mc^{2}$ caused him to
think about both dynamics of planets/satellites and light in the gravitational
field of heavy mass on equal footing [31]. Very likely, this equal footing
requirement was sufficient for developing of his relativity theory without
consulting \ the works by Poincare$^{^{\prime}}$on celestial mechanics \ in
which dynamics of light was not discussed. Analogously, for Heisenberg the
main question was: \ To what extent can one restore the underlying microscopic
dynamic system using combinatorial analysis of the observed spectral data?
\ Surprisingly, the full answer \ to this question compatible with
Heisenberg's original ideas had been obtained only quite recently. Details and
references can be found in our work, Ref.[15]. For the sake of space, in this
paper we only provide absolute minimum of results needed for correct modern
understanding of Heisenberg's ideas.

We begin with observation that the Schr\"{o}dinger equation cannot be reduced
to something else which is related to our macroscopic experience. \textsl{It
has to} \textsl{be postulated}.\footnote{Usually used appeal to the DeBroigle
wave-particle duality is of no help since the wave function in the
Schr\"{o}dinger's equation plays an auxiliary role.} On the contrary,
Heisenberg's basic equation from which all quantum mechanics can be recovered
is directly connected with experimental data and looks almost trivial. Indeed,
following Bohr, Heisenberg looked at the famous equations for energy levels
difference%
\begin{equation}
\omega(n,n-\alpha)=\frac{1}{\hbar}(E(n)-E(n-\alpha)), \tag{2.6}%
\end{equation}
where both $n$ and $n-\alpha$ are some integers. He noticed [26] that this
definition leads to the following fundamental composition law:%

\begin{equation}
\omega(n-\beta,n-\alpha-\beta)+\omega(n,n-\beta)=\omega(n,n-\alpha-\beta).
\tag{2.7a}%
\end{equation}
Since by design $\omega(k,n)=-\omega(n,k),$ the above equation can be
rewritten in a symmetric form as
\begin{equation}
\omega(n,m)+\omega(m,k)+\omega(k,n)=0. \tag{2.7b}%
\end{equation}
In such a form it is known as the honeycomb equation (condition) in current
mathematics literature [12-14] where it was rediscovered totally independently
of \ Heisenberg's \ key quantum mechanical paper\ and, apparently, with
different purposes in mind. Connections between \ mathematical results of
Knutson and Tao [12-14] and those of Heisenberg were noticed and discussed in
recent paper by Kholodenko[15,16]. We would like to use some results from this
work now.

We begin by noticing that (2.7b) due to its purely combinatorial origin does
not contain the Plank's constant $\hbar$. Such fact is of major importance for
this work since the condition (2.4) can be equivalently rewritten in the form
of (2.7b), where $\omega(n,m)=\omega_{n}-\omega_{m}$. It would be quite
unnatural to think of the Planck's constant in this case\footnote{Planck's
constant is normally being used for objects at the atomic scales interacting
with light. However, there are systems other than atomic, e.g. polymers, in
which Schr\"{o}dinger (or even Dirac-type])-type equations \ are being used
with Planck's constant being replaced by the stiffness parameter-different for
different polymers [32].Incidentally, conformational properties of very stiff
(helix-type) polymers are described by the neutrino-type equation, etc.
[33].}. Equation (2.7b) is essentially of the same type as (2.4). It looks
almost trivial and yet, it is sufficient for restoration of all quantum
mechanics. Indeed, in his paper of October 7th of 1925, \ Dirac[34], \ being
aware of Heisenberg's key paper\footnote{This paper was sent to Dirac by
Heisenberg himself.}, streamlined Heisenberg's results and \ introduced
\ notations which are in use up to this day. He noticed that the combinatorial
law given by (2.7a) for frequencies, when used in the Fourier expansions for
composition of observables, leads to the multiplication rule
$a(nm)b(mk)=ab(nk)$ for the Fourier amplitudes for these observables. In
general, in accord with Heisenberg's assumptions, one expects that $ab(nk)\neq
ba(nk).$ Such a multiplication rule is typical for matrices. In the modern
quantum mechanical language such matrix elements are written as $<n\mid\hat
{O}\mid m>\exp(i\omega(n,m)t)$ so that (2.7.b) is equivalent to the matrix
statement%
\begin{align}%
%TCIMACRO{\tsum \nolimits_{m}}%
%BeginExpansion
{\textstyle\sum\nolimits_{m}}
%EndExpansion
&  <n\mid\hat{O}_{1}\mid m><m\mid\hat{O}_{2}\mid k>\exp(i\omega(n,m)t)\exp
(i\omega(m,k)t)\nonumber\\
&  =<n\mid\hat{O}_{1}\hat{O}_{2}\mid k>\exp(i\omega(n,k)t). \tag{2.8}%
\end{align}
for some operator (observables) $\hat{O}_{1}$ and $\hat{O}_{2}$ evolving
according to the rule: $\hat{O}_{k}(t)=U\hat{O}_{k}U^{-1},k=1,2,$ provided
that $U^{-1}=\exp(-i\frac{\hat{H}}{\hbar}t).$ \ From here it follows that
$U^{-1}\mid m>=\exp(-\frac{E_{m}}{\hbar}t)\mid m>$ \ if one identifies
$\hat{H}$ with the Hamiltonian operator. Clearly, upon such an identification
the Schr\"{o}dinger equation can be obtained at once as is well known [35] and
with it, the rest of quantum mechanics. In view of Ref.s[12-16] it is possible
to extend the traditional pathway: from classical to quantum mechanics and
back. This topic is discussed in the next section.

\section{Space, time and space-time in classical and quantum mechanics}

\subsection{General comments}

If one contemplates quantization of dynamics of celestial objects using
traditional textbook prescriptions, one will immediately run into myriad of
small and large problems. Unlike atomic systems in which all electrons repel
each other, have the same masses and \ are indistinguishable, in the case of,
say, Solar System all planets (and satellites) attract each other, have
different masses and visibly distinguishable. Besides, in the case of atomic
systems the Planck constant $\hbar$ plays prominent role while no such a role
can be given to the Planck constant in the sky.

In the previous section it was demonstrated that in the Einsteinian limit it
is possible to remove the above objections so that, apparently, the only
difference between the atomic and celestial quantum mechanics lies in
\ replacement of the Planck constant by another constant to be determined in
Section 4.

\subsection{Space and time in classical and quantum mechanics}

Although \ celestial mechanics based on Newton's law of gravity \ is
considered to be classical (i.e. non quantum), \ with such an assumption one
easily runs into serious problems. Indeed, such an assumption \ implies that
the speed with which the interaction propagates is infinite and that time is
the same everywhere. Wether this is true or false can be decided only
experimentally. Since at the scales of our Solar System one has to use radio
signals to check correctness of Newton's celestial mechanics, one is faced
immediately with all kind of wave mechanics effects such as retardation, the
Doppler effect, etc. Because of this, measurements are necessarily having some
error margins. The error margins naturally will be larger for more distant
objects. \textsl{Accordingly, even at the level of classical mechanics applied
to} \textsl{the motion of celestial bodies we have to deal with certain
inaccuracies similar in nature to those} \textsl{in atomic mechanics.} To make
formalisms of both atomic and celestial quantum mechanics look the same we
have to think carefully about the space, time and space-time transformations
already at the level of classical mechanics.

We begin with observation that in traditional precursor of quantum
mechanics-the Hamiltonian mechanics-the Hamiltonian equations \textsl{by
design} remain invariant with respect to the canonical transformations. That
is if \ sets $\{q_{i}\}$ and \{$p_{i}\}$ represent the "old" canonical
coordinates and momenta while $Q_{i}=Q_{i}(\{q_{i}\},\{p_{i}\})$ and
$P_{i}=P_{i}(\{q_{i}\},\{p_{i}\})$, $i=1-N$, represent the "new" set of
canonical coordinates and momenta, the Hamiltonian equations in the old
variables given by
\begin{equation}
\dot{q}_{i}=\frac{\partial H}{\partial p_{i}}\text{ \ and }\dot{p}_{i}%
=-\frac{\partial H}{\partial q_{i}} \tag{3.1}%
\end{equation}
and those rewritten in "new" \ variables will have the same form. Here we used
the commonly accepted notations, e.g. $\dot{q}_{i}=\frac{d}{dt}q_{i}$ , etc.
Quantum mechanics uses this form-invariance essentially as is well known. \ 

We would like to complicate this traditional picture by investigating the
"canonical " time changes in classical mechanics. Fortunately, such task was
accomplished to a large extent in the monograph by Pars [36]. For the sake of
space, we refer our readers to pages 535-540 of this monograph for more
details. Following Dirac [37]$,$ we notice that good quantization procedure
should always begin with the Lagrangian formulation of mechanics since it is
not always possible to make a transition from the Lagrangian to Hamiltonian
form of mechanics (and, thus, to quantum mechanics) due to presence of some
essential constraints (typical for mechanics of gauge fields, etc.). Hence, we
also begin with the Lagrangian functional $\mathcal{L=L(}\{q_{i}\},\{\dot
{q}_{i}\})$. The Lagrangian equations of motion can be written in the form of
Newton's equations $\dot{p}_{i}=F_{i},$ where the generalized momenta $p_{i}$
are given by $p_{i}=\delta\mathcal{L}/\delta\dot{q}_{i}$ and the generalized
forces $F_{i}$ by $F_{i}=-\delta\mathcal{L}/\delta q_{i}$ as usual. In the
case if the total energy $E$ is conserved, it is possible instead of "real"
time $t$ to introduce the fictitious time $\theta$ via relation $dt=u(\{q_{i}%
\})d\theta$ where the function $u(\{q_{i}\})$ is assumed to be nonnegative and
is sufficiently differentiable with respect to its arguments. At this point we
can enquire if Newton's equations can be written in terms of new time variable
so that they remain form- invariant. To do so, following Pars, we must: \ a)
to replace $\mathcal{L}$ by $u\mathcal{L},$ b) \ to replace $\dot{q}_{i}$ by
$\ q_{i}^{\prime}$ $/u,$ where $q_{i}^{\prime}$=$\frac{d}{d\theta}q_{i}$, c)
to rewrite new Lagrangian in terms of \ such defined new time variables and,
finally, d) to obtain Newton's equations according to the described rules,
provided that now we have to use $p_{i}^{\prime}$ instead of $\dot{p}_{i}$. In
the case if the total energy of the system is conserved, we shall obtain back
the same form of Newton's equations rewritten in terms of new variables. This
means that by going from the Lagrangian to Hamiltonian formalism of classical
mechanics we can write the Hamilton's equations (3.1) in which the dotted
variables are replaced by primed. Furthermore, (3.1) will remain the same if
we replace the Hamiltonian $H$ by some nonnegative function $f(H)$ while
changing time $t$ to time $\theta$ according to the rule $d\theta
/dt=df(H)/dH\mid_{H=E}$. Such a change while leaving classical mechanics
form-invariant will affect quantum mechanics where now the Schr\"{o}dinger's
equation%
\begin{equation}
i\hbar\frac{\partial}{\partial t}\Psi=\hat{H}\Psi\tag{3.2}%
\end{equation}
is replaced by
\begin{equation}
i\hbar\frac{\partial}{\partial\theta}\Psi=f(\hat{H})\Psi. \tag{3.3}%
\end{equation}
With such information at our hands, we would like to discuss the extent to
which symmetries of our (empty) space-time affect dynamics of particles
"living" in it.

\subsection{Space-time in quantum mechanics}

\subsubsection{\bigskip General comments}

Use of group-theoretic methods in quantum mechanics had began almost
immediately after its birth. It was initiated by Pauli in 1926. He obtained a
complete quantum mechanical solution for\ the Hydrogen atom \ employing
symmetry arguments. His efforts were not left without appreciation. Our
readers can find many historically important references in two comprehensive
review papers by Bander and Itzykson [38]. In this subsection we pose and
solve the following problem: Provided that the symmetry of (classical or
quantum) system is known, will this information be sufficient for
determination of this system uniquely?
%TCIMACRO{\TEXTsymbol{\backslash}}%
%BeginExpansion
$\backslash$%
%EndExpansion

Below, we shall provide simple and concrete examples illustrating meaning of
the word "determination". In the case of quantum mechanics this problem is
known as the problem about hearing of the "shape of the drum". It was
formulated by Mark Kac [39]. The problem can be formulated as follows. Suppose
that the sound spectrum of the drum is known, will such an information
determine the shape of the drum uniquely? The answer is "No" [40]. Our readers
may argue at this point that non uniqueness could come as result of our
incomplete knowledge of symmetry or, may be, as result of the actual lack of
true symmetry (e.g. the Jahn-Teller effect in molecules, etc. in the \ case of
quantum \ mechanics). \ These factors do play some role but they cannot be
considered as decisive as the basic example below demonstrates.

\subsubsection{ Difficulties with the correspondence principle for Hydrogen
atom}

In this subsection we do not use arguments by Kac \ since our arguments are
much more straightforward. We choose the most studied case of Hydrogen atom as
an example.

As it is well known, the Keplerian motion of a particle in the centrally
symmetric field is planar \ and is exactly solvable for both the scattering
and bound states at the classical level [36]\textbf{. }The result of such a
solution depends on two parameters: the energy and the\textbf{\ }%
angular\textbf{\ }momentum. The correspondence principle formulated by Bohr is
expected to provide the bridge between the classical and quantum realities by
requiring that in the limit of large quantum numbers the results of quantum
and classical calculations for observables should coincide. However, this
requirement may or may not be possible to implement. It is violated already
for the Hydrogen atom! Indeed, according to the naive canonical quantization
prescriptions, one should begin with the \textsl{classical} Hamiltonian in
which one has to replace the momenta and coordinates by their operator
analogs. Next, one uses such constructed quantum Hamiltonian in the
Schr\"{o}dinger's equation, etc. Such a procedure breaks down at once for the
Hamiltonian of Hydrogen atom since the intrinsic planarity of the classical
\ Kepler's problem is entirely ignored thus leaving the projection of the
angular momentum without its classical analog. Accordingly, the
\textsl{scattering differential crossection for Hydrogen atom obtained quantum
mechanically (within} \textsl{the 1st Born approximation) uses essentially
3-dimensional formalism and coincides} \textsl{with the classical result by
Rutherford obtained for planar configurations!} Thus, even for the Hydrogen
atom classical and quantum (or, better, pre quantum) Hamiltonians\textbf{\ }%
\textsl{do not} match thus formally violating the correspondence principle.
Evidently, semiclassically we can only think of energy and the angular
momentum thus leaving the angular momentum projection undetermined. Such a
"sacrifice" is justified by the agreement between the observed and predicted
Hydrogen atom spectra and by use of Hydrogen-like atomic orbitals for
multielectron atoms, etc. Although, to our knowledge, such a mismatch is not
mentioned in any of the students textbooks on quantum mechanics, its existence
is essential if we are interested in extension of quantum mechanical ideas to
dynamics of Solar System. In view of such an interest, we would like to
reconsider traditional treatments of Hydrogen atom, this time being guided
\ only by the symmetry considerations. This is accomplished in the next subsection.

\subsubsection{Emergence of the SO(2,1) symmetry group}

In April of 1940 Jauch and Hill\textbf{\ [}41\textbf{]} published a paper in
which they studied the\ planar Kepler problem quantum mechanically. Their work
was stimulated by earlier works by Fock of 1935 and by Bargmann of 1936 in
which it was shown that the spectrum of bound states for the Hydrogen atom can
be obtained by using representation theory of SO(4) group of rigid rotations
of 4-dimensional Euclidean space while the spectrum of scattering states can
be obtained by using the Lorentzian group SO(3,1). By adopting results of Fock
and Bargmann to the planar configuration Jauch and Hill obtained the
anticipated result: In the planar case one should use SO(3) group for the
bound states and SO(2,1) group for the scattering states. Although \ this
result will be reconsidered almost entirely, we mention about it now having
several purposes in mind.

First, we would like to reverse arguments leading to the final results of
Jauch and Hill in order to return to the problem posed at the beginning of
this section. That is, we want to use the fact that the Kepler problem is
planar (due to central symmetry of the force field) and the fact that the
motion takes place in (locally) Lorentzian space-time in order to argue that
the theory of group representations for Lorentzian SO(2,1) symmetry
group-intrinsic for this Kepler problem- \ correctly reproduces the Jauch-Hill
spectrum. Nevertheless, the question remains: Is Kepler's problem the only one
exactly solvable classical and quantum mechanical problem associated with the
SO(2,1) group? Below we demonstrate that this is not the case! In
\ anticipation of such negative result, we would like to develop our intuition
by using some known results from quantum mechanics.

\subsubsection{\bigskip Classical-quantum correspondence allowed by SO(2,1)
symmetry: a gentle introduction}

For the sake of space, we consider here only the most generic (for this work)
example in some detail: the radial Schr\"{o}dinger equation for the planar
Kepler problem with the Coulombic potential. It is given by\footnote{The
rationale for discussing the Coulombic potential instead of gravitational will
be fully explained in the next section.}%
\begin{equation}
-\frac{\hbar^{2}}{2\mu}(\frac{d^{2}}{d\rho^{2}}+\frac{1}{\rho}\frac{d}{d\rho
}-\frac{m^{2}}{\rho^{2}})\Psi(\rho)-\frac{Ze^{2}}{\rho}=E\Psi(\rho). \tag{3.4}%
\end{equation}
Here $\left\vert m\right\vert =0,1,2,...$ is the angular momentum quantum
number as required. For $E<0$ it is convenient to introduce the dimensionless
variable $x$ via $\rho=ax$ and to introduce the new wave function: $\psi
(\rho)=\sqrt{\rho}\Psi(\rho)$. Next, by the appropriate choice of constant $a$
and by redefining $\psi(\rho)$ as $\psi(\rho)=\gamma x^{\frac{1}{2}+\left\vert
m\right\vert }\exp(-y)\varphi(y),$ where $y=\gamma x,$ -$\gamma^{2}=\frac{2\mu
E}{\hbar^{2}}a^{2},a=\frac{\hbar^{2}}{\mu ZE},$ the following hypergeometric
equation can be eventually obtained:%
\begin{equation}
\left\{  y\frac{d^{2}}{dy^{2}}+2[\left\vert m\right\vert +\frac{1}{2}%
-y]\frac{d}{dy}+2[\frac{1}{\gamma}-\left\vert m\right\vert -\frac{1}%
{2}]\right\}  \varphi(y)=0. \tag{3.5}%
\end{equation}
Formal solution of such an equation can be written \ as $\varphi
(y)=\mathcal{F}(-A(m),B(m),y),$ where $\mathcal{F}$ is the confluent
hypergeometric function. Physical requirements imposed on this function reduce
it to a polynomial leading to the spectrum of the planar Kepler problem.
Furthermore, by looking into standard textbooks on quantum mechanics, one can
easily find that \textsl{exactly the same type of hypergeometric equation} is
obtained for problems such as one-dimensional Schr\"{o}dinger's equation with
the Morse-type potential,\footnote{That is, $V(x)=A(exp(-2\alpha
x)-2exp(-\alpha x)).$} three dimensional radial Schr\"{o}dinger equation for
the harmonic oscillator\footnote{That is, $V(r)=\dfrac{A}{r^{2}}+Br^{2}.$} and
even three dimensional radial equation for the Hydrogen atom\footnote{That is,
$V(r)=\dfrac{A}{r^{2}}-\dfrac{B}{r}.$}. Since the two-dimensional Kepler
problem is solvable with help of representations of SO(2,1) \ Lorentz group,
the same should be true for all quantum problems just listed. That this is the
case is demonstrated, for example, in the book by Wybourne [42]. A sketch of
the proof is provided in Appendix A. This proof indicates that, actually, the
\textsl{discrete spectrum} of all problems just listed is obtainable with help
of SO(2,1) group. The question remains: If the method outlined in Appendix A
provides the spectra of several quantum mechanical problems listed above, can
we be sure that these are the only exactly solvable quantum mechanical
problems associated with the SO(2,1) Lorentz group? \ Unfortunately, the
answer is "No"! More details are given below.

\subsubsection{\bigskip Common properties of quantum mechanical problems
related to SO(2,1) Lorentz group}

In Appendix A a sketch of the so called spectrum-generating algebras (SGA)
method is provided. It is aimed at producing the exactly solvable one-variable
quantum mechanical problems. In this subsection we would like to put these
results in a broader perspective. In particular, in our works[15,16] we
demonstrated that\textsl{\ all exactly} \textsl{solvable quantum mechanical
problem should involve hypergeometric functions of single or multiple
arguments}. We argued that the difference between different problems can be
understood topologically in view of the known relationship between
hypergeometric functions and braid groups. These results, even though quite
rigorous, are not well adapted for immediate practical use. In this regard
more useful would be to solve the following problem: \textsl{For a given}
\textsl{set of orthogonal polynomials find the corresponding many-body
operator for which such a set of orthogonal polynomials forms the complete set
of} \textsl{eigenfunctions}. At the level of orthogonal polynomials of one
variable relevant for all exactly solvable two-body problems of quantum
mechanics, one can think about the related problem of finding all potentials
in one-dimensional radial Schr\"{o}dinger equation, e.g. equation (A.1),
leading to the hypergeometric-type solutions. Very fortunately, such a task
was accomplished already by Natanzon [43]. Subsequently, his results were re
investigated by many authors with help of different methods, including SGA. To
our knowledge, the most complete recent summary of the results, including
potentials and spectra can be found in the paper by Levai [44]. Even this
(very comprehensive) paper does not cover all aspects of the problem. For
instance, it does not mention the fact that these results had been extended to
relativistic equations such as Dirac and Klein-Gordon for which similar
analysis was made by Cordero with collaborators [45]. In all cited cases
(relativistic and non relativistic) the underlying symmetry group was SO(2,1).
The results of Appendix A as well as of all other already listed references
can be traced back to the classically written papers by Bargmann [46] and
Barut and Fronsdal [47] on representations of SO(2,1) Lorentz group.
Furthermore, the discovered connection \ of this problematic with
supersymmetric quantum mechanics [48,49] can be traced back to the 19th
century works by Gaston Darboux. The fact that representations of the
\textsl{planar} SO(2,1) Lorentz group are sufficient to cover all known
exactly solvable two-body problems (instead of the full SO(3,1) Lorentz
group!) is quite remarkable. It is also sufficient for accomplishing the
purposes of this work-to quantize the dynamics of Solar System- but leaves
open the question : Will use of the full Lorentz group lead to the exactly
solvable quantum mechanical problems not accounted by the SO(2,1) group
symmetry? \ This topic will be briefly discussed in Section $\mathbf{5}$. In
the meantime, we would like to address the problem of quantizing the Solar
System dynamics using \ the obtained results This is accomplished in the next section.

\section{\ Quantum celestial mechanics of Solar System}

\subsection{ General remarks}

We begin this subsection by returning back to (2.4). Based on previous
discussions, this equation provides us with opportunity to think seriously
about quantum nature of dynamics of our Solar System dynamics. Nevertheless,
such an equation reveals only one aspect of quantization and, as such,
provides only sufficient condition for quantization. The necessary condition
in atomic and celestial mechanics lies in the \textsl{non dissipativity} of
dynamical systems in both cases. Recall that Bohr introduced his quantization
prescription to avoid dissipation caused by the emission of radiation \ by
electrons in orbits in general position. New quantum mechanics have
\textsl{not} \ shed much light on absence of dissipation for stationary Bohr's
orbits. At the level of old Bohr theory absence of dissipation at the
stationary Bohr orbit was explained by Boyer [50]. Subsequently, his result
was refined by Puthoff\textbf{\ [}51\textbf{]}. In the case of Solar System
absence of dissipation for motion on stable orbits was discussed by Goldreich
[52] who conjectured that the \ dissipative (tidal) effects adjust the initial
motion of planets/satellites in such a way that eventually the orbits become
stable. More on this is discussed in Subsection 5.2. Notice that, dynamics of
Solar System\ as considered by Poincare$^{^{\prime}}$ and by those who
developed his ideas does not involve treatment of tidal effects! Treatment of
tidal effects in general relativity represents one of the \ serious challenges
for this theory [53]. Thus, very much by analogy with Bohr, \textsl{we have to
postulate} that in the case of Solar System (Hamiltonian) dynamics of stable
orbits is non dissipative. This assumption then leads us to the following Table:\ 

\ \ \ \ \ \ \ \ \ \ \ \ \ \ \ \ \ \ \ \ \ \ \ \ \ \ \ \ \ \ \ \ \ \ \ \ \ \ \ \ \ \ \ \
\[
\text{Table 1}%
\]
\ \ \ \ \ \
\begin{tabular}
[t]{|l|l|l|}\hline%
\begin{tabular}
[c]{l}%
%TCIMACRO{\TEXTsymbol{\backslash}}%
%BeginExpansion
$\backslash$%
%EndExpansion
$Type$\textit{\ }$of$\textit{\ }$mechanics$\\
$\mathit{Properties}$%
\end{tabular}
&
\begin{tabular}
[c]{l}%
$Quantum$\textit{\ }$atomic$\\
$mechanics$%
\end{tabular}
&
\begin{tabular}
[c]{l}%
$\mathit{Quantum}$\\
$celestial$\textit{\ }$mechanics$%
\end{tabular}
\\\hline%
\begin{tabular}
[c]{l}%
Dissipation (type of)%
%TCIMACRO{\TEXTsymbol{\backslash}}%
%BeginExpansion
$\backslash$%
%EndExpansion
\\
(yes%
%TCIMACRO{\TEXTsymbol{\backslash}}%
%BeginExpansion
$\backslash$%
%EndExpansion
no)%
%TCIMACRO{\TEXTsymbol{\backslash}}%
%BeginExpansion
$\backslash$%
%EndExpansion
on stable orbits
\end{tabular}
&
\begin{tabular}
[c]{l}%
electromagnetic\\
friction%
%TCIMACRO{\TEXTsymbol{\backslash}}%
%BeginExpansion
$\backslash$%
%EndExpansion
no%
%TCIMACRO{\TEXTsymbol{\backslash}}%
%BeginExpansion
$\backslash$%
%EndExpansion
\\
Bohr orbits
\end{tabular}
&
\begin{tabular}
[c]{l}%
tidal friction\\%
%TCIMACRO{\TEXTsymbol{\backslash}}%
%BeginExpansion
$\backslash$%
%EndExpansion
no%
%TCIMACRO{\TEXTsymbol{\backslash}}%
%BeginExpansion
$\backslash$%
%EndExpansion
Einstein's geodesics
\end{tabular}
\\\hline%
\begin{tabular}
[c]{l}%
Accidental degeneracy%
%TCIMACRO{\TEXTsymbol{\backslash}}%
%BeginExpansion
$\backslash$%
%EndExpansion
\\
(yes%
%TCIMACRO{\TEXTsymbol{\backslash}}%
%BeginExpansion
$\backslash$%
%EndExpansion
no)%
%TCIMACRO{\TEXTsymbol{\backslash}}%
%BeginExpansion
$\backslash$%
%EndExpansion
origin
\end{tabular}
& yes%
%TCIMACRO{\TEXTsymbol{\backslash}}%
%BeginExpansion
$\backslash$%
%EndExpansion
Bohr-Sommerfeld condition & yes%
%TCIMACRO{\TEXTsymbol{\backslash}}%
%BeginExpansion
$\backslash$%
%EndExpansion
Laplace condition\\\hline
Charge neutrality & yes & no(but see below)\\\hline
Masses &
\begin{tabular}
[c]{l}%
electrons are having\\
the same masses
\end{tabular}
&
\begin{tabular}
[c]{l}%
(up to validity of the\\
equivalence principle)\\
masses are the same
\end{tabular}
\\\hline
Minimal symmetry group & SO(2,1) & SO(2,1)\\\hline
Correspondence principle & occasionally violated & occasionally
violated\\\hline%
\begin{tabular}
[c]{l}%
Discrete spectrum:\\
finite or infinite%
%TCIMACRO{\TEXTsymbol{\backslash}}%
%BeginExpansion
$\backslash$%
%EndExpansion
reason%
%TCIMACRO{\TEXTsymbol{\backslash}}%
%BeginExpansion
$\backslash$%
%EndExpansion
\\
Pauli principle(yes%
%TCIMACRO{\TEXTsymbol{\backslash}}%
%BeginExpansion
$\backslash$%
%EndExpansion
no)
\end{tabular}
&
\begin{tabular}
[c]{l}%
finite and infinite%
%TCIMACRO{\TEXTsymbol{\backslash}}%
%BeginExpansion
$\backslash$%
%EndExpansion
\\
charge neutrality%
%TCIMACRO{\TEXTsymbol{\backslash}}%
%BeginExpansion
$\backslash$%
%EndExpansion
\\
yes
\end{tabular}
&
\begin{tabular}
[c]{l}%
finite%
%TCIMACRO{\TEXTsymbol{\backslash}}%
%BeginExpansion
$\backslash$%
%EndExpansion
\\
no charge neutrality%
%TCIMACRO{\TEXTsymbol{\backslash}}%
%BeginExpansion
$\backslash$%
%EndExpansion
\\
yes
\end{tabular}
\\\hline
\end{tabular}

\bigskip

\subsubsection{\bigskip Celestial spectroscopy and the Titius-Bode law of
planetary distances}

The atomic spectroscopy was inaugurated by Newton in the second half of 17th
century. The celestial spectroscopy was inaugurated by Titius in the second
half of 18th century and become \ more famous after it was advertised by
Johann Bode, the Editor of the "Berlin Astronomical Year-book". The book by
Nieto [54] provides extensive bibliography related to uses and interpretations
of the Titius-Bode (T-B) law up to second half of 20th century. Unlike the
atomic spectroscopy, where the observed atomic and molecular spectra were
expressed using simple empirical formulas which were (to our knowledge) never
elevated to the status of \ "law", in celestial mechanics the empirical T-B
formula
\begin{equation}
r_{n}=0.4+03.\cdot2^{n}\text{, \ \ \ }n=-\infty,0,1,2,3,... \tag{4.1}%
\end{equation}
for the orbital radii (semimajor axes) of planets acquired the status of a law
in the following sense. In the case of atomic spectroscopy the empirical
formulas used for description of atomic/molecular spectra have not been used
(to our knowledge) for making predictions. Their purpose was just to describe
in mathematical terms what had been already observed. Since the T-B empirical
formula for planetary distances was used as the law, it was used in search for
planets not yet discovered. In such a way Ceres, Uranus, Neptune and Pluto
were found [10]. However, the discrepancies for Neptune and Pluto were much
larger than the error margins allowed by the T-B law. This fact divided the
astronomical community into "believers" and "atheists"\ (or non believers)
regarding to the meaning and uses of this law. Without going into historical
details, we would like to jump to the very end of the Titius-Bode story in
order to use its latest version \ which we found in the paper by Neslu\v{s}an
[55] who, in turn, was motivated by the work of Lynch [56]. Instead of (4.1)
these authors use another empirical power law dependence
\begin{equation}
r_{n}=r_{0}B^{n},\text{ }n=1,2,3,..,9. \tag{4.2}%
\end{equation}
For planets (except Pluto and including the asteroid belt) Neslu\v{s}an
obtained\footnote{In astronomical units (to be defined below).} $r_{0}%
(au)=0.203$ and $B=1.773$ with the rms deviation accuracy of
0.0534\footnote{This result gives for the Earth in astronomical (au) units the
result $r_{3}\simeq1.13.$ Much better result is obtained in case if we choose
$B=1.7.$ In this case we obtain: $r_{3}\simeq.997339.$ Lynch provides
$B=1.706$ and $r_{0}=0.2139.$}. Analogous power law dependencies were obtained
previously in the work by Dermott [57] for both planets and satellites of
heavy planets such as Jupiter, Saturn and Uranus.

It should be noted that \ because of noticed discrepancies the attempts were
made to prove or disprove the Titius-Bode law by using statistical analysis,
e.g. see papers by Lynch [56] and Hayes and Tremaine [58], with purpose of
finding out to which extent the observed dependencies can be considered as non
accidental. Following logic of Bohr we would like to use the observed
empirical radial dependencies as a guide to our calculations to be discussed
below. We leave with astronomers to resolve the semantic aspects of these
observed dependencies.

\subsubsection{An attempt at quantization of celestial(Solar System) dynamics}

Being guided by the Table 1 at the beginning of this section we will be
assuming that planets do not interact since they move along geodesics
independently. In the case of atomic mechanics it was clear from the beginning
that such an approximation should sooner or later fail. The nonexisting
electroneutrality in the sky \ provides strong hint that the T-B law must be
of very limited use since the number of discrete levels \ for gravitating
systems should be \textsl{always finite}. Otherwise, we would observe the
countable infinity of satellites around \ Sun or of any of heavy planets. This
is \ not observed and is physically wrong. It is wrong because such a system
would tend to capture all matter in the Universe.

In the literature one can find many attempts at quantization of Solar System
using standard prescriptions of quantum mechanics. Since this work is not a
review, we do not provide references to papers whose results do not affect
ours\footnote{With one exception to be mentioned in Section 5.}. \textsl{Blind
uses of standard rules of quantum mechanics \ for quantization of our Solar
System do not contain any provisions \ for finite number of energy
levels/orbits for gravitating systems. }

To facilitate matters in the present case, we would like to make several
additional observations. First, we have to find an analog of the Planck
constant. Second, we have to have some mechanical model in mind to make our
search for physically correct answer successful. To accomplish the first task,
we have to take into account the 3-rd Kepler's law. In accord with (2.3), it
can be written as $r_{n}^{3}/T_{n}^{2}=\dfrac{4\pi^{2}}{\gamma(M_{\odot}+m)}$.
In view of arguments presented in the Subsection 2.1 , we can safely
approximate this result by $4\pi^{2}/\gamma M_{\odot}$, where $M_{\odot}$ is
mass of the Sun. For the \ purposes of this work, it is convenient to restate
this law as $3lnr_{n}-2\ln T_{n}=\ln4\pi^{2}/\gamma M_{\odot}=const$
\ \ Below, we choose the \textsl{astronomical system of units} in which
$4\pi^{2}/\gamma M_{\odot}=1.$ By definition, in this system of units we have
for the Earth: $r_{3}=T_{3}=1$.

Consider now the Bohr result (2.6) and take into account that $E=\hbar
\omega\equiv\dfrac{h}{2\pi}\dfrac{2\pi}{T}.$ Therefore, Bohr's result can be
conveniently restated as $\omega(n,m)=\omega(n)-\omega(m).$ Taking into
account \ equations (2.6),(3.3),(4.2) and the third Kepler's law, we formally
obtain:%
\begin{equation}
\omega(n,m)=\frac{1}{c\ln\tilde{A}}(nc\ln\tilde{A}-mc\ln\tilde{A}), \tag{4.3}%
\end{equation}
where the role of Planck's constant is played now by $c\ln\tilde{A}$ ,
$\tilde{A}=B^{\frac{3}{2}}$ and $c$ is some constant which will be determined
selfconsistently below\footnote{Not to be confused with the speed of light
!}$.$

At first, one may think that what we obtained is just a simple harmonic
oscillator spectrum. After all, this should come as not too big a surprise
since in terms of the action -angle variables all exactly integrable systems
are reducible to the sets of harmonic oscillators. This result is also
compatible with the results of Appendix A. The harmonic oscillator option is
physically undesirable in the present case though since the harmonic
oscillator has countable infinity of energy levels. Evidently, such a spectrum
is equivalent to the T-B law. But it is well known that this law is not
working well for larger numbers. In fact, it would be very strange should it
be working in this regime in view of arguments already presented.

To make a progress, we have to use the 3rd Kepler's law once again, i.e. we
have to take into account that in the astronomical system of units $3lnr_{n}=$
$2\ln T_{n}.$ A quick look at equations (A.11), (A.12) suggests that the
underlying mechanical system is likely to be associated with that for the
Morse potential. \ This is so because the low lying states of such a system
cannot be distinguished from those for the harmonic oscillator. However, this
system does have only a finite number of energy levels which makes sense
physically. The task remains to connect this system with the planar Kepler's
problem. Although in view of results of Appendix A such a connection does
indeed exist, we want to demonstrate it explicitly at the level of classical mechanics.

Before doing so we have to make several comments. First, according to the
Table 1, the planets/satellites should move along the geodesics. Second, the
geodesics which were used by Einstein for bending of light and for motion of
Mercury are obtainable with help of the metric coming from the Schwarzshild's
\ solution of Einstein's equations for pure gravity in the vacuum [31].
Clearly, one can think about uses of Kerr solutions, Weyl solutions, etc. as
well for the same purposes. Such thinking is perfectly permissible but is not
of much help for a particular case of our Solar System. In it, Kepler's laws
describe reality sufficiently well so that Einsteinian geodesics can be safely
approximated by the Newtonian orbits. This conclusion is perfectly compatible
with the original Einstein's derivation of his equations for gravity. Thus, we
are adopting his strategy in our paper. Clearly, once the results are
obtained, they can be recalculated if needed. Fortunately, as far as we can
see, there is no need for doing this as we shall demonstrate shortly below.

Following Pars [36], the motion of a point of unit mass in the field of
\ Newtonian gravity is described by the following equation%
\begin{equation}
\dot{r}^{2}=(2Er^{2}+2\gamma Mr-\alpha^{2})/r^{2}, \tag{4.4}%
\end{equation}
where $\alpha$ is the angular momentum integral (e.g. see equation (5.2.55) of
Pars book). We would like now to replace $r(t)$ by $r(\theta)$ in such a way
that $dt=u(r$($\theta)$)$d\theta$ . Let therefore $r(\theta)=r_{0}%
\exp(x(\theta)),$ -$\infty<x<\infty.$ Unless otherwise specified, we shall
write $r_{0}=1$. In such (astronomical) system of units) we obtain, $\dot
{r}=x^{\prime}\dfrac{d\theta}{dt}\exp(x(\theta)).$ This result can be further
simplified by choosing $\dfrac{d\theta}{dt}=\exp(-x(\theta)).$ With this
choice (4.4) acquires the following form:%
\begin{equation}
(x^{\prime})^{2}=2E+2\gamma M\exp(-x)-\alpha^{2}\exp(-2x). \tag{4.5}%
\end{equation}
Consider points of equilibria for the potential \ $U(r)=-2\gamma
Mr^{-1}+\alpha^{2}r^{-2}.$ Using it, we obtain: $r^{\ast}=\dfrac{\alpha^{2}%
}{\gamma M}.$ According to Goldstein et al\ [20] such defined $r^{\ast}$
coincides with the major elliptic semiaxis. It can be also shown, e.g. Pars,
equation (5.4.14), that for the Kepler problem the following relation holds:
$\ E=-\dfrac{\gamma M}{2r^{\ast}}$. Accordingly, $r^{\ast}=-\dfrac{\gamma
M}{2E},$ and, furthermore, using the condition $\frac{dU}{dr}=0$ we obtain:
$\dfrac{\alpha^{2}}{\gamma M}=-\dfrac{\gamma M}{2E}$ or, $\alpha^{2}$
$=-\dfrac{\left(  \gamma M\right)  ^{2}}{2E}.$ Since in the chosen system of
units $r(\theta)=\exp(x(\theta)),$ we obtain as well: $\dfrac{\alpha^{2}%
}{\gamma M}=\exp(x^{\ast}(\theta)).$ It is convenient to choose $x^{\ast
}(\theta)=0.$ This requirement makes the point $x^{\ast}(\theta)=0$ as the
origin and implies that with respect to such chosen origin $\alpha^{2}=\gamma
M.$ In doing so some caution should be exercised since upon quantization
equation $r^{\ast}=\dfrac{\alpha^{2}}{\gamma M}$ becomes $r_{n}^{\ast}%
=\dfrac{\alpha_{n}^{2}}{\gamma M}.$ By selecting the astronomical scale
$r_{3}^{\ast}=1$ as the unit of length implies then that we can write the
angular momentum $\alpha_{n}^{2}$ as $\varkappa$ $\dfrac{r_{n}^{\ast}}%
{r_{3}^{\ast}}$ and \ to define $\varkappa$ as $\alpha_{3}^{2}$ $\equiv
\alpha^{2}.$ Using this fact (4.5) can then be conveniently rewritten as
\begin{equation}
\frac{1}{2}(x^{\prime})^{2}-\gamma M(\exp(-x)-\frac{1}{2}\exp(-2x))=E
\tag{4.6a}%
\end{equation}
or, equivalently, as
\begin{equation}
\frac{p^{2}}{2}+A(\exp(-2x)-2\exp(-x))=E, \tag{4.6b}%
\end{equation}
where $A=\dfrac{\gamma M}{2}.$ \ \ Since this result is exact classical analog
of the quantum Morse potential problem, transition to quantum mechanics can be
done straightforwardly at this stage. By doing so \ we have to replace the
Planck's constant $\hbar$ by $c\ln\tilde{A}$. After that, we can write the
answer for the spectrum at once [59]:%
\begin{equation}
-\tilde{E}_{n}=\frac{\gamma M}{2}[1-\frac{c\ln\tilde{A}}{\sqrt{\gamma M}%
}(n+\frac{1}{2})]^{2}. \tag{4.7}%
\end{equation}

This result contains an unknown parameter $c$ to be determined now. To do so
it is sufficient to expand the potential in (4.6b) and to keep terms up to
quadratic. Such a procedure produces the anticipated harmonic oscillator
result%
\begin{equation}
\frac{p^{2}}{2}+Ax^{2}=\tilde{E} \tag{4.8}%
\end{equation}
with the quantum spectrum given by $\tilde{E}_{n}=(n+\frac{1}{2})c\sqrt{2A}%
\ln\tilde{A}.$ In the astronomical system of units the spectrum reads:
$\tilde{E}_{n}=(n+\frac{1}{2})c2\pi\ln\tilde{A}$ . This result is in agreement
with (4.3). To proceed, we \ notice that in (4.3) the actual sign of the
Planck-type constant is undetermined. Specifically, in our case (up to a
constant) the energy $\tilde{E}_{n}$ is determined by ln$\left(  \frac
{1}{T_{n}}\right)  $ $=-\ln$ $\tilde{A}$ so that it makes sense to write
$-\tilde{E}_{n}\sim n\ln\tilde{A}.$ To relate the classical energy defined by
the Kepler-type equation $E=-\dfrac{\gamma M}{2r^{\ast}}$ to the energy we
just have defined, we have to replace the Kepler-type equation by $-\tilde
{E}_{n}\equiv-\ln\left\vert E\right\vert =-2\ln\sqrt{2}\pi+\ln r_{n}\text{
\ }$This is done in view of the 3rd Kepler's law and the fact that the new
coordinate $x$ is related to the old coordinate $r$ via $r=e^{x}$. Using (4.2)
(for $n=1$) in the previous equation and comparing it with the already
obtained spectrum of the harmonic oscillator we obtain:
\begin{equation}
-2\ln\sqrt{2}\pi+\ln r_{0}B=-c2\pi\ln\tilde{A}, \tag{4.9}%
\end{equation}
where in arriving at this result we had subtracted the nonphysical ground
state energy. Thus, we obtain:%
\begin{equation}
c=\frac{1}{2\pi\ln\tilde{A}}\ln\frac{2\pi^{2}}{r_{0}B}. \tag{4.10}%
\end{equation}
Substitution of this result back into (4.7) produces%
\begin{align}
-\tilde{E}_{n}  &  =2\pi^{2}[1-\frac{(n+\frac{1}{2})}{4\pi^{2}}\ln\left(
\frac{2\pi^{2}}{r_{0}B}\right)  ]^{2}\simeq2\pi^{2}[1-\frac{1}{9.87}%
(n+\frac{1}{2})]^{2}\nonumber\\
&  \simeq2\pi^{2}-4(n+\frac{1}{2})+0.2(n+\frac{1}{2})^{2}. \tag{4.11}%
\end{align}
To determine the number of bound states, we follow the same procedure as was
developed long ago in chemistry for the Morse potential. \ For this
purpose\footnote{Recall, that in chemistry the Morse potential is being
routinely used for description of the vibrational spectra of diatomic
molecules.} we introduce the energy difference $\Delta\tilde{E}_{n}=$
$\tilde{E}_{n+1}-\tilde{E}_{n}=4-0.4(n+1)$ first. Next, the maximum number of
bound states is determined by requiring $\Delta\tilde{E}_{n}=0.$ In our case,
we obtain: $n_{\max}=9$. This number is in perfect accord with observable data
for planets of our Solar System (with Pluto being excluded and the asteroid
belt included). In spite of such a good accord, some caution must be still
exercised while analyzing the obtained result. Should we not insist on
physical grounds that the discrete spectrum must contain only finite number of
levels, the obtained spectrum for the harmonic oscillator would be sufficient
(that is to say, that the validity of the T-B law would be confirmed).
Formally, it \ also solves the quantization problem completely and even is in
accord with the numerical data [55]. The problem lies however in the fact that
these data were fitted to the power law (4.2) in accord with the original T-B
empirical guess. Heisenberg's honeycomb rule (2.7b) does \textsl{not} rely on
specific $n-$dependence. \ In fact, we have to consider the observed (the
Titius-Bode-type) $n-$dependence only as a hint, especially because in this
work we intentionally avoid use of any adjustable parameters. The developed
procedure, when supplied with correctly interpreted numerical data, is
sufficient for obtaining results without any adjustable parameters as we just
demonstrated. In turn, this allows to replace the T-B law in which the power
$n$ is unrestricted by more accurate result working especially well for larger
values of $n$. \ For instance, the constant $c$ was determined using the
harmonic approximation for the Morse-type potential. This approximation is
expected to fail very quickly as the following arguments indicate. Although
$r_{n}^{\prime}s$ can calculated using the T-B law given by (4.2), the
arguments following this equation cause us to look also at the equation
$-\tilde{E}_{n}\equiv-\ln\left\vert E\right\vert =-2\ln\sqrt{2}\pi+\ln
r_{n\text{ \ }}$ for this purpose. This means that we have to use (4.11) (with
ground state energy subtracted) in this equation in order to obtain the result
for $r_{n}.$ If we ignore the quadratic correction in (4.11) (which is
equivalent of calculating the constant $c$ using harmonic oscillator
approximation to the Morse potential) then, by construction, we recover the
T-B result (4.2). If, however, we do not resort to such an approximation,
calculations will become much more elaborate. The final result will indeed
replace the T-B law but its analytical form is going to be too cumbersome for
practicioners. Since corrections to the harmonic oscillator potential in the
case of the Morse potential are typically small, they do not change things
qualitatively. Hence, we do not account for these complications in our paper.
Nevertheless, accounting for these (anharmonic) corrections readily explains
why the empirical T-B law works well for small n's and becomes increasingly
unreliable for larger n's [54].

In support of our way of \ doing quantum calculations, we would like to
discuss now similar calculations for satellite systems of Jupiter, Saturn,
Uranus and Neptune. \ To do such calculations the astronomical system of units
is not immediately useful since in the case of heavy planets one cannot use
the relation $4\pi^{2}/\gamma M_{\odot}=1.$ This is so because we have to
replace the mass of the Sun $M_{\odot}$ by the mass of respective heavy
planet. For this purpose we write $4\pi^{2}=$\ $\gamma M_{\odot}$, multiply
both \ sides by $M_{j\text{ }}$\ (where $j$ stands for the $j$-th heavy
planet) and divide both sides by $M_{\odot}$. Thus, we obtain: $4\pi^{2}%
q_{j}=$\ $\gamma M_{j}$,\ where $q_{j}=$\ $\dfrac{M_{j}}{M_{\odot}}$\ . Since
the number $q_{j}$ is of order $10^{-3}$\ $-10^{-5}$, it causes some
inconveniences in actual calculations. To avoid this difficulty, we need to
readjust (4.6a)\ by rescaling $x$ coordinate as $x=\delta\bar{x}$\ and, by
choosing \ $\delta^{2}$\ $=q_{j}$. After transition to quantum mechanics such
a rescaling results in \ replacing (4.7) for the spectrum by the following
result:%
\begin{equation}
-\tilde{E}_{n}=\frac{\gamma M}{2}[1-\frac{c\delta\ln\tilde{A}}{\sqrt{\gamma
M}}(n+\frac{1}{2})]^{2}. \tag{4.12}%
\end{equation}
Since the constant $c$\ is initially undetermined, we can replace it by
$\tilde{c}=c\delta.$ This replacement \ allows us to reobtain back equation
almost identical to (4.11). That is
\begin{equation}
-\tilde{E}_{n}=2\pi^{2}[1-\frac{(n+\frac{1}{2})}{4\pi^{2}}\ln\left(
\frac{\gamma M_{j}}{(r_{j})_{1}}\right)  ]^{2} \tag{4.13}%
\end{equation}
In this equation \ $\gamma M_{j}=$\ $4\pi^{2}q_{j}$ and \ $(r_{j})_{1}$ is the
semimajor axis of the satellite lying in the equatorial plane and closest to
the $j$-th planet. Our calculations are summarized in the Table 2 below.
Appendix B contains the input data used in calculations of n$_{theory}^{\ast
}.$\ Observational data are taken from the web link given in the 1st footnote.\ \ \ \ \ \ \ \ \ \ \ \ \ \ \ \ \ \ \ \ \ \ \ \ \ \ \ \ \ \ \ \ \ \ \ 

\ \ \ \ \ \ \ \ \ \ \ \ \ \ \ \ \ \ \ \ \ \ \ \ \ \ \ \ \ \ \ \ \ \ \ \ \ \ \ \ \ Table
2%
\[%
\begin{tabular}
[c]{|l|l|l|}\hline
Satellite system%
%TCIMACRO{\TEXTsymbol{\backslash}}%
%BeginExpansion
$\backslash$%
%EndExpansion
n$_{\max}$ & n$_{theory}^{\ast}$ & n$_{\text{obs}}^{\ast}$\\\hline
Solar system & 9 & 9\\\hline
Jupiter system & 11-12 & 8\\\hline
Saturn system & 20 & 20\\\hline
Uranus system & 40 & 18\\\hline
Neptune system & 33 & 6-7\\\hline
\end{tabular}
\ \
\]
Since the discrepancies for Uranus and Neptune systems may be genuine or not
we come up with the following general filling pattern which is being compared
with that discussed in the Introduction.

\subsection{Filling patters in Solar System: similarities and differences with
atomic mechanics}

From atomic mechanics we know that the approximation of independent electrons
used by Bohr fails rather quickly with increased number of electrons.
\ Already for this reason to expect that the T-B law is going to hold for
satellites of heavy planets is naive. At the same time, for planets rotating
around the Sun such an approximation is seemingly good but also not without
flaws. The SO(2,1) symmetry explains why motion of all planets should be
planar but it does not explain why motion of all planets is taking place in
the plane coinciding with the equatorial plane of the Sun or why all planets
are moving in the same direction. The same is true for the regular satellites
of all heavy planets as discussed by Dermott [57]. Such a configuration can be
explained by a plausible hypothesis [7] that all planets of Solar System and
regular satellites of heavy planes are originated from evolution of the
pancake-like cloud. This assumption is not without problems though. For
instance, all irregular satellites and Saturn's Phoebe ring are rotating in
the "wrong" direction. Under conditions of such a hypothesis all these objects
were randomly captured by the already existing Solar System at later times.
The exoplanets rotating in the wrong directions\footnote{E.g.read the
Introduction} apparently also had been captured so that the origins of some
planetary systems are quite different from that for ours if we believe that
ours originated from the pancake- like cloud. This hypothesis would make sense
should irregular satellites be arranged around respective planets at random.
But they are not! This is discussed in the Introduction. In view of these
facts, in this work we tend to provide a quantum mechanical explanation of the
observed filling patterns summarized in Table 1. This Table requires some
extension, for instance, to account for the fact that all planets and regular
satellites are moving in the respective equatorial planes. \ This fact can be
accounted for by the effects of spin-orbital interactions. Surprisingly, these
effects exist both at the classical, newtonian, level [11] and at the level of
general relativity [60].\ \ At the classical level the most famous example of
spin-orbital resonance (but of a different kind) is exhibited by the motion of
the Moon whose orbital period coincides with its rotational period so that it
always keeps only one face towards the Earth. Most of the major natural
satellites are locked in analogous 1:1 spin-orbit resonance with respect to
the planets around which they rotate. Mercury represents an exception since it
is locked into 3:2 resonance around the Sun (that is Mercury completes 3/2
rotation around its axis while making one full rotation along its orbit).
Goldreich [52]\ explains such resonances as results of influence of
dissipative (tidal) processes on evolutionary dynamics of Solar System. The
resonance structures observed in the sky are stable equilibria in the
appropriately chosen reference frames [11].\ \ Clearly, the spin-orbital
resonances just described\ are not explaining many things. For instance, while
nicely explaining why our Moon is always facing us with the same side the same
pattern is not observed for Earth rotating around the Sun with exception of
Mercury. Mercury is treated as pointlike object in general relativity.
Decisive attempt to describe the motion of extended objects in general
relativity was made in seminal paper by Papapetrou [61] and continues up to
the present day.\ From his papers it is known that, strictly speaking, motion
of the extended bodies is not taking place on geodescs. And yet, for the
Mercury such an approximation made originally by Einstein works extremely
nicely.\ The spin-orbital interaction [60] causing planets and satellites of
heavy planets to lie in the equatorial plane is different from that causing
\ 1:1, etc. resonances. It is analogous to the NMR-type resonances in atoms
and molecules where in the simplest case we are dealing, say, with the
Hydrogen atom. \ In it, the proton having spin 1/2 \ is affected by the
magnetic field created by the \ orbital s-electron. In the atomic case due to
symmetry of electron s-orbital this effect is negligible but nonzero! This
effect is known in chemical literature as "chemical shift".\ In celestial case
the situation is similar but the effect is expected to be much stronger since
the orbit is planar (not spherical as for the s-electron in hydrogen atom,
e.g. see Subsection 3.3.2).\ Hence, the equatorial location of planetary
orbits and regular satellites is likely the result of such spin-orbital
interaction. The equatorial plane in which planets (satellites) move can be
considered as some kind of an orbital (in terminology of atomic physics).\ It
is being filled in accordance with the equivalent of the Pauli
principle:\textsl{\ each orbit can be occupied by no more than one
planet}\footnote{The meteorite belt can be looked upon as some kind of a ring.
We shall discuss the rings below, in the next subsection.}. Once the orbital
is filled, other orbitals (planes) will begin to be filled out. Incidentally,
such a requirement automatically excludes Pluto from the status of a planet.
Indeed, on one hand, the T-B-type law, can easily accommodate Pluto, on
another, not only this would contradict the data summarized in Table 1 and the
results of previous subsection but also, and more importantly, \ it would be
in contradiction with the astronomical data for Pluto. According to these data
the orbit inclination for Pluto is 17$^{o}$ as compared to the rest of planets
whose inclination is within boundary margins of $\pm$2$^{o}($ except for
Mercury for which it is 7$^{o})$. Some of the orbitals can be empty and
\textsl{not all orbits belonging to the same orbital (a plane) must be filled}
(\textsl{as it is also the case in atomic physics}) . This is indeed observed
in the sky[10,62] and is consistent with results of \ Table 2. It should be
said though that it appears (according to available data\footnote{E.g. see
footnote 1 and [10].}, that not all of the observed satellites are moving on
stable orbits. It appears\ also as if and when the "inner shell"is completely
filled, it acts as some kind of an s-type spherical orbital since \textsl{the}
\textsl{orbits of} \textsl{other (irregular) satellites lie strictly outside
the sphere whose diameter is greater or equal to that corresponding to the
last allowed energy level in the first shell}. \ In accord with results of
previous subsection, the location of secondary planes appears to be quite
arbitrary as well as filling of their stable orbits. Furthermore, \ without
account of spin-orbital interactions, one can say nothing about the direction
of orbital rotation. Evidently, the "chemical shift" created by the orbits of
regular satellites lying in the s-shell is such that it should be more
energetically advantageous to rotate in the opposite direction. This
proposition requires further study. In addition to planets and satellites on
stable orbits there are many strangers in the Solar System: comets,
meteorites, etc. These are moving not on stable orbits and, as result, should
either leave the Solar System or \ eventually collide with those which move on
"legitimate" orbits.

\ It is tempting to extend the picture just sketched beyond the scope of our
Solar System. If for a moment we would ignore relativistic effects (they will
be discussed in the next section), we can then find out that our Sun is moving
along almost circular orbit around our galaxy center with the period
$T=185\cdot10^{6}$ years [63]. Our galaxy is also flat as our Solar System and
the major mass is concentrated in the galaxy center. Hence, again, \ if we
believe that stable stellar motion \ is taking place along the geodesics
around the galaxy center in accord with laws of Einstein's general relativity,
then \ we have to accept that our galaxy is also a quantum object. It would be
very interesting to estimate the number of allowed energy levels (stable
orbits) for our galaxy and to check if the Pauli-like principle works for our
and other galaxies as well.\ 

\subsection{The restricted 3-body problem and planetary rings}

\ \ \ \ \ \ \ Although the literature on the restricted 3-body problem is
huge,\ we would like to discuss this problem from the point of view of \ its
connection with general relativity and quantization of planetary orbits
\ along the lines advocated in this paper. We begin with several remarks.
First, the existence of ring systems for \textsl{all} heavy planets is well
documented [10]. Second, these ring systems are interspersed with satellites
of these planets. Third, both rings and satellites lie in the respective
equatorial planes (with exception of Phoebe's ring) so that satellites move on
stable orbits. From these observations it follows that: \ \ \ \ 

a) While each of heavy planets is moving along the geodesics around the Sun,
the respective satellites are moving along the geodesics around \ respective planets;

b) \ The motion of these satellites is almost circular (the condition which
Laplace took into account while studying Jupiter's \ regular satellites).

\textsl{The restricted 3-body problem can be formulated now as follows}.

Given that the rings are made of some kind of small objects whose masses can
be neglected\footnote{This approximation is known as Hill's
problem/approximation in the restricted 3-body problem [25, 36].} as compared
to masses of both satellite(s) and the respective heavy planet, we can ignore
mutual gravitational \ interaction between these objects (as Laplace did).
Under such conditions we end up with the motion of a given piece of a ring (of
zero mass) in the presence of two bodies of masses $m_{1}$ and $m_{2}$
respectively (the planet and one of the regular satellites). To simplify
matters, it is usually being assumed that the motion of these two masses takes
place on a\ circular orbit with respect to their center of mass. Complications
associated with the eccentricity of such a motion are discussed in the book by
Szebehely [64] and can be taken into account if needed. They will be ignored
nevertheless in our discussion since we shall assume that satellites of heavy
planets move on geodesics so that the center of mass coincides with the
position of a heavy planet anyway thus making our computational scheme
compatible with Einsteinian relativity. By assuming that ring pieces are
massless we also are making their motion compatible with requirements of
general relativity, \ since whatever orbits they may have-these are geodesics anyway.\ \ \ \ \ \ \ \ \ \ \ \ \ \ \ \ \ \ \ \ \ \ \ \ \ \ \ \ \ \ \ \ \ \ \ \ \ \ \ \ \ \ \ \ \ \ \ \ \ \ \ \ \ \ \ \ \ \ \ \ \ \ \ \ \ \ \ \ \ \ \ \ \ \ \ \ \ \ \ \ \ \ \ \ \ \ \ \ \ \ \ \ \ \ \ \ \ \ \ \ \ \ \ \ \ \ \ \ \ \ \ \ \ \ \ \ \ \ \ \ \ \ \ \ \ \ \ 

Thus far only the motion of \ regular satellites in the equatorial planes (of
respective planets) was considered as stable (and, hence, quantizable). The
motion of ring pieces was not accounted by these stable orbits. The task now
lies in showing that satellites \textsl{lying inside} \textsl{the respective
rings} \textsl{of heavy planets \ are essential for stability of these rings
motion thus making it} \textsl{quantizable}. For the sake of space, we would
like only to provide a sketch of arguments leading to such a conclusion. Our
task is greatly simplified by the fact that very similar situation exists for
3-body system such as Moon, Earth and Sun. \ Dynamics of such a system was
studied thoroughly by Hill whose work played pivotal role in Poincare$^{\prime
}$ studies of celestial mechanics \ [6]. Avron and Simon\textbf{\ [}%
65\textbf{]} adopted Hill's ideas in order to develop formal quantum
mechanical treatment of the Saturn rings. In this work we follow instead the
original Hill's ideas of dynamics of the Earth-Moon-Sun system. When these
ideas are looked upon from the point of view of modern mathematics of exactly
integrable systems, they enable us to describe not only the Earth-Moon-Sun
system but also the dynamics of rings of heavy planets. These modern
mathematical methods allow us to find a place for the Hill's theory within
general quantization scheme discussed in previous sections.

\subsubsection{\bigskip Basics of the Hill's equation}

To avoid repetitions, we refer our readers to the books of Pars [36] and
Chebotarev [63] for detailed and clear account of the restricted 3-body
problem and Hill's contributions to Lunar theory. Here we only summarize the
ideas behind Hill's ground breaking work.

In a nutshell his method of studying the Lunar problem can be considered as
extremely sophisticated improvement of previously mentioned Laplace method.
Unlike Laplace, Hill realized that \ both Sun and Earth are surrounded by the
rings of influence\footnote{Related to the so called Roche limit [10,36].}.
The same goes for all heavy planets. Each of these planets and each satellite
of such a planet will have its own domain of influence whose actual width is
controlled by the Jacobi integral of motion. For the sake of argument,
consider the Saturn as an example. It has Pan as its the innermost satellite.
Both the Saturn and Pan have their \ respective domains of influence.
Naturally, we have to look first at the domain of influence for the Saturn.
Within such a domain let us consider a hypothetical closed Kepler-like
trajectory. \ Stability of such a trajectory is described by the Hill
equation\footnote{In fact, there will be the system of Hill's equations in
\ general [63]. This is so since the disturbance of trajectory is normally
decomposed into that which is perpendicular and that \ which is parallel to
the \ Kepler's trajectory at a given point. We shall avoid these complications
in our work.}. Since such an equation describes a wavy-type oscillations
around the presumably stable trajectory, the parameters describing such a
trajectory are used as an input (perhaps, with subsequent adjustment) in the
Hill equation given by%
\begin{equation}
\frac{d^{2}x}{dt^{2}}+(q_{0}+2q_{1}\cos2t+2q_{2}\cos4t+\cdot\cdot
\cdot)x=0.\tag{4.14}%
\end{equation}
If we would ignore all terms except $q_{0}$ first, we would \ naively obtain:
$x_{0}(t)=A_{0}cos$($t\sqrt{q_{0}}+\varepsilon).$ This result describes
oscillations around the equilibrium position along the trajectory with the
constant $q_{0}$ carrying information about this trajectory. The amplitude $A$
is expected to be larger or equal to the average distance between the pieces
of the ring. This naive picture gets very complicated at once should we use
the obtained result as an input into (4.14). In this case the following
equation is obtained
\begin{equation}
\frac{d^{2}x}{dt^{2}}+q_{0}x+A_{0}q_{1}\{\cos[t(\sqrt{q_{0}}+2)+\varepsilon
]+\cos[t(\sqrt{q_{0}}-2)-\varepsilon]\}=0\tag{4.15}%
\end{equation}
whose solution will enable us to determine $q_{1}$and $A_{1}$ using the
appropriate boundary conditions. Unfortunately, since such a procedure should
be repeated \ infinitely many times, it is obviously impractical. Hill was
able to design a much better method. \ Before discussing Hill's equation from
the perspective of modern mathematics, it is useful \ to recall the very basic
classical facts \ about this equation summarized in the book by Ince [66]. For
this purpose, we shall assume that the solution of (4.14) can be presented in
the form
\begin{equation}
x(t)=e^{\alpha t}%
%TCIMACRO{\tsum \limits_{r=-\infty}^{\infty}}%
%BeginExpansion
{\textstyle\sum\limits_{r=-\infty}^{\infty}}
%EndExpansion
b_{r}e^{irt}.\tag{4.16}%
\end{equation}
Substitution of this result into (4.14) leads to the following infinite system
of linear equations%
\begin{equation}
(\alpha+2ri)^{2}b_{r}+%
%TCIMACRO{\tsum \limits_{k=-\infty}^{\infty}}%
%BeginExpansion
{\textstyle\sum\limits_{k=-\infty}^{\infty}}
%EndExpansion
q_{k}b_{r-k}=0,\text{ }r\in\mathbf{Z.}\tag{4.17}%
\end{equation}
As in finite case, obtaining of the nontrivial solution requires the infinite
determinant $\Delta(\alpha)$ to be equal to zero. \ This problem can be looked
upon from two directions: either all constants $q_{k}$ are assigned and one is
interested in the bounded solution of (4.16) for $t\rightarrow\infty$ or, one
is interested in the relationship between constants \ made in such a way that
$\alpha=0.$ In the last case it is important to know wether there is one or
more than one of such solutions available. Although answers can be found in
the book by Magnus and Winkler [67], we follow McKean and Moerbeke [68],
Trubowitz [69] and Moser [70].

For this purpose, we need to bring our notations in accord with those used in
these references. Thus, the Hill operator is defined now as $Q(q)=-\frac
{d^{2}}{dt^{2}}+q(t)$ with periodic potential $q(t)=q(t+1).$ Equation (4.14)
can now be rewritten as
\begin{equation}
Q(q)x=\lambda x. \tag{4.18}%
\end{equation}
This representation makes sense since $q_{0}$ in (4.14 ) plays the role of
$\lambda$ in $(4.18).$ Since this is the second order differential equation,
it has formally 2 solutions. These solutions depend upon the boundary
conditions. For instance, for \textsl{periodic} solutions such that
$x(t)=x(t+2)$ the "spectrum" of (4.18) is discrete and is given by
\[
-\infty<\lambda_{0}<\lambda_{1}\leq\lambda_{2}<\lambda_{3}\leq\lambda
_{4}<\cdot\cdot\cdot\uparrow+\infty.
\]
We wrote the word spectrum in quotation marks because of the following.
Equation (4.18) does have a normalizable solution only if $\lambda$ belongs to
the (pre assigned) intervals $(\lambda_{0},\lambda_{1}),(\lambda_{2}%
,\lambda_{3}),...,(\lambda_{2i},\lambda_{2i+1}),...$ \ \ In such a case the
eigenfunctions $x_{i}$ are normalizable in the usual sense of quantum
mechanics and form the orthogonal set. Periodic solutions make sense only for
vertical displacement from the reference trajectory. For the horizontal
displacement the boundary condition should be chosen as $x(0)=x(1)=0.$ For
such chosen boundary condition the discrete spectrum also exists but it lies
exactly in the gaps between the intervals just described, i.e. $\lambda
_{1}\leq\mu_{1}\leq\lambda_{2}<\lambda_{3}\leq\mu_{2}\leq\lambda_{4}%
<\cdot\cdot\cdot.$ For such a spectrum there is also set of normalized
mutually orthogonal eigenfunctions. \textsl{Thus in both cases quantum
mechanical} \textsl{description is assured}. One can do much more however. In
particular, Trubowitz [69] designed an explicit procedure for recovering the
potential $q(t)$ from the $\mu-$spectrum supplemented by information about
normalization constants.

\ \ It is quite remarkable that the Hill's equation can be interpreted in
terms of the auxiliary dynamical (Neumann) problem. Such an interpretation is
very helpful for us since it allows us to include the quantum mechanics of
Hill's equation into general formalism developed in this work.

\subsubsection{Connection with the dynamical Neumann problem and the Korteweg
-de Vries equation}

Before describing such connections, we would like to add few details to the
results of previous subsection. First, as in the planetary case, the number of
pre assigned intervals is always finite. This means that, beginning with some
pre assigned $\hat{\imath}$, we would be left with $\lambda_{2i}%
=\lambda_{2i+1}\forall i>\hat{\imath}.$ These \textsl{double} eigenvalues do
not have independent physical significance since they can be determined by the
set of \textsl{single} eigenvalues (for which $\lambda_{2i}\neq\lambda
_{2i+1})$ as demonstrated by Hochstadt [71]. Because of this, potentials
$q(t)$ in the Hill's equation are called the \textsl{finite gap}
potentials\footnote{Since there is only finite number of gaps [$\lambda
_{1},\lambda_{2}],$[$\lambda_{3},\lambda_{4}],...$where the spectrum is
forbidden.}. Hence, physically, it is sufficient to discuss only potentials
which possess finite single spectrum. The auxiliary $\mu-$spectrum is then
determined by the gaps of the single spectrum as explained above. With this
information in our hands, we are ready to discuss the exactly solvable Neumann
dynamical problem. It is the problem about dynamics of a particle moving on
$n-$dimensional sphere $<\mathbf{\xi},\mathbf{\xi}>\equiv\xi_{1}^{2}$
+$\cdot\cdot\cdot+\xi_{n}^{2}=1$ under the influence of a quadratic potential
$\phi(\mathbf{\xi})=<\mathbf{\xi},\mathbf{A\xi}>.$ Equations of motion
describing the motion on $n-$ sphere are given by
\begin{equation}
\mathbf{\ddot{\xi}}=-\mathbf{A\xi}+u(\mathbf{\xi})\mathbf{\xi}\text{ \ with
}u(\mathbf{\xi})=\phi(\mathbf{\xi})-<\mathbf{\dot{\xi}},\mathbf{\dot{\xi}%
}>.\tag{4.19}%
\end{equation}
Without loss of generality, we assume that the matrix $\mathbf{A}$ is already
in the diagonal form: $\mathbf{A}:=diag(\alpha_{1},...,\alpha_{n}).$ With such
an assumption we can equivalently rewrite Eq.(4.19) in the following
suggestive form%
\begin{equation}
\left(  -\frac{d^{2}}{dt^{2}}+u(\mathbf{\xi}(t))\right)  \xi_{k}=\alpha_{k}%
\xi_{k}\text{ ; \ }k=1,...,n.\tag{4.20}%
\end{equation}
Thus, in the case if we can prove that $u(\mathbf{\xi}(t))$ in (4.19) is the
same as $q(t)$ in (4.18), the connection between the Hill and Neumann's
problems will be established. The proof is presented in Appendix C. It is
different from that given in the lectures by Moser [70] since it is more
direct and much shorter.

This proof brought us the unexpected connection with hydrodynamics through
\ the static version of Korteweg-de Vries equation. Attempts to describe the
Saturnian rings using equations of hydrodynamic are described in the recent
monograph by Esposito [72]. This time, however, we can accomplish more using
\ just obtained information. This is the subject of the next subsection.

\subsubsection{Connections with SO(2,1) group and the K-Z equations}

\bigskip

Following Kirillov [73], we introduce the commutator for the fields
(operators) $\xi$ and $\eta$ as follows: $[\xi,\eta]=\xi\partial\eta
-\eta\partial\xi.$ Using the KdV equation (C.10), let us consider 3 of its
independent solutions: $\xi_{0},\xi_{-1}$ and $\xi_{1}.$ All these solutions
can be obtained from general result: $\xi_{k}=t^{k+1}+O(t^{2}),$ valid near
zero. Consider now a commutator $[\xi_{0},\xi_{1}].$ Straightforwardly, we
obtain, $[\xi_{0},\xi_{1}]=\xi_{1}$. Analogously, we obtain, $[\xi_{0}%
,\xi_{-1}]=-\xi_{-1}$ and, finally, $[\xi_{1},\xi_{-1}]=-2\xi_{0}.$ According
to Kirillov, such a Lie algebra is isomorphic to that for the group $SL(2,R)$
which is the center for the Virasoro algebra\footnote{Since connections
between the KdV and the Virasoro algebra are well documented [74], it is
possible in principle to reinterpret fine structure of the Saturn's rings
string-theoretically.}. Vilenkin [75] demonstrated that the group $SL(2,R)$ is
isomorphic to $SU(1,1)$. Indeed, by means of transformation: \textit{w}%
\textsl{=}$\dfrac{\mathit{z}-i}{\mathit{z}+i},$ it is possible to transform
the upper half plane (on which $SL(2,R)$ acts) into the interior of unit
circle on which $SU(1,1)$ acts. Since, according to Appendix A, the group
$SU(1,1)$ is the connected component of $SO(2,1)$, the anticipated connection
with $SO(2,1)$ group is established.

In Appendix C we noticed connections between the Picard-Fuchs, Hill and
Neumann-type equations. In a recent paper by Veselov et al [76] such a
connection was developed much further resulting in the Knizhnik-Zamolodchikov-
type equations for the Neumann-type dynamical systems. We refer our readers to
original literature, especially to the well written lecture notes by Moser
[70]. These notes as well and his notes in collaboration with Zehnder [30]
provide an excellent background for study the whole circle of ideas ranging
from Hill's equation and integrable models to string theory,etc.

\section{Solar System at larger scales: de Sitter, anti -de Sitter and
conformal symmetries compatible with orbital quantization}

Results obtained in previous section demonstrate remarkable interplay between
the Newtonian and Einsteinian mechanics at the scale of our Solar System.
Quantization of stable (Laplace-Einstein) orbits makes sense only with
\ account of observational/empirical facts unequivocally supporting  general
relativity. It is only natural to reverse this statement and to say
that\textsl{ the observed \ filling patterns of stable (quantum) orbits is yet
another manifestation  of general relativity}.

Since quantum mechanics can be developed group-theoretically, the same should
be true for relativity. Quoting Einstein, Infel'd and Hoffmann [77]:
"Actually, the only equations of gravitation which follow without ambiguity
from the fundamental assumptions of the general theory of relativity are the
equations for empty space, and it is important to know whether they alone are
capable of determining the motion of bodies". \ The results of this work
strongly support such a conclusion. Given this, we would like to discuss how
such locally Lorentzian space-time embeds into space-times of general
relativity possessing larger symmetry groups. Since this topic is extremely
large, we shall discuss only the most basic facts from the point of view of
results obtained in this paper.

To our knowledge, Dirac [78] was the first who recognized the role of
space-time symmetry in quantum mechanics. In his paper he wrote: "The
equations of atomic physics are usually formulated in terms of space-time of
special relativity. They then have to form a scheme which remains invariant
under all transformations which carry the space-time over into itself. These
transformations consist of the Lorentz rotations about a point combined with
arbitrary translations, and form a group.... Nearly all of more general spaces
have only trivial groups\footnote{This statement of Dirac is not correct.
However, it was correct \ based on mathematical knowledge at the time of
writing of his paper.}of operations which carry the spaces into
themselves....There is one exception, however, namely the de Sitter space
(with no local gravitational fields). This space is associated with a very
interesting group, and so the study of the equations of atomic physics in this
space is of special interest, \textsl{from mathematical point of view}."
Subsequent studies indicated that the symmetry of space-time could be
important even at the atomic scale [79,80]. This fact suggests that quantum
mechanics of Solar System can be potentially useful for studies in cosmology,
e.g. for studies of the cosmological constant problem [79,81], of cold dark
energy (CDE) [82], of cold dark matter (CDM) [83] and of the modified
Newtonian dynamics (MOND) [84]. Clearly, we are unable to discuss these issues
within the scope of this paper. Nevertheless, we would like to notice that,
for instance, the MOND presupposes use of Newtonian and the modified Newtonian
mechanics at the galactic scales which, strictly speaking, \ is not
permissible. As we had argued, it is not permissible even at the scales of our
Solar System. Mathematical rationale behind what is called in literature as
"dark energy and dark matter" is explained in our recent paper [85]. In it we
discussed some physical applications of mathematical results by Grisha
Perelman used in his proof of the Poincare$^{\prime}$ and geometrization conjectures.

As by-product of \ the results discussed in [85], we \ would like to discuss
briefly a simple construction of the de Sitter and anti-de Sitter spaces. We
begin with the \ Hilbert-Einstein functional
\begin{equation}
S^{c}(g)=%
%TCIMACRO{\tint \nolimits_{\mathcal{M}}}%
%BeginExpansion
{\textstyle\int\nolimits_{\mathcal{M}}}
%EndExpansion
d^{d}xR\sqrt{g}+\Lambda%
%TCIMACRO{\tint \nolimits_{\mathcal{M}}}%
%BeginExpansion
{\textstyle\int\nolimits_{\mathcal{M}}}
%EndExpansion
d^{d}x\sqrt{g}\tag{5.1}%
\end{equation}
defined for some (pseudo) Riemannian manifold $\mathcal{M}$ of total
space-time dimension $d$. The (cosmological) constant $\Lambda$ \ is just the
Lagrangian multiplier assuring volume conservation. It is determined as
follows\footnote{It should be noted though that mathematicians study related
but not identical problem of minimization of the Yamabe functional, given by
$Y(g)=\left(
%TCIMACRO{\tint \nolimits_{\mathcal{M}}}%
%BeginExpansion
{\textstyle\int\nolimits_{\mathcal{M}}}
%EndExpansion
d^{d}xR\sqrt{g}\right)  /$ $\left(
%TCIMACRO{\tint \nolimits_{\mathcal{M}}}%
%BeginExpansion
{\textstyle\int\nolimits_{\mathcal{M}}}
%EndExpansion
d^{d}x\sqrt{g}\right)  ^{\frac{2}{p}}$ with $p=2d/(2-d)$, e.g. see our papers
[86]. It is conformal invariant -different for different manifolds. Only at
the mean field level results of minimization of $S^{c}(g)$ coincide with those
obtainable by minimization of $Y(g).$ In this work this approximation is
sufficient.}. \ On one hand, with help of the Ricci curvature tensor $R_{ij}$,
the \textsl{Einstein space} is defined as solution of the equation%
\begin{equation}
R_{ij}=\lambda g_{ij}\tag{5.2}%
\end{equation}
with $\lambda$ being a constant. From this definition it follows that
$R=d\lambda.$ On another hand, variation of the action $S^{c}(g)$ produces%
\begin{equation}
G_{ij}+\frac{1}{2}\Lambda g_{ij}=0,\tag{5.3}%
\end{equation}
where the Einstein tensor $G_{ij}=R_{ij}-\frac{1}{2}g_{ij}R$ with $R$ being
the scalar curvature determined by the metric tensor $g_{ij}\footnote{Eq.(5.3)
illustrates the meaning of the term "dark matter". The constant $\Lambda$
enters into the stress-energy tensor (typically associated with matter)..In
the present case it is given by $-$ $\frac{1}{2}\Lambda g_{ij})$.}.$ The
combined use of (5.2) and (5.3) produces: $\Lambda=\lambda(d-2).$ Substitution
of this result back into (5.3) produces:%
\begin{equation}
G_{j}^{i}=(\frac{1}{d}-\frac{1}{2})\delta_{j}^{i}R.\tag{5.4}%
\end{equation}
Since\ by design $G_j,i^i=0,$ \ we obtain our major result:
\begin{equation}
(\frac{1}{d}-\frac{1}{2})R_{,j}=0,\tag{5.5}%
\end{equation}
implying that the scalar curvature $R$ \ for Einstenian spaces is a constant.

For isotropic homogenous spaces the Riemann curvature tensor can be presented
in the form [87]:
\begin{equation}
R_{ijkl}=k(x)(g_{ik}g_{jl}-g_{il}g_{jk}).\tag{5.6}%
\end{equation}
Accordingly, the Ricci tensor is obtained as: $R_{ij}=k(x)g_{ij}(d-1).$
Schur's theorem [87] guarantees that for $d\geq3$ we must have $k(x)=k=const$
for the entire space. Therefore, we obtain: $\lambda=(d-1)k$ and, furthermore,
$R=d(d-1)k.$ The spatial coordinates can always be rescaled so that $R=k$ or,
alternatively, the constant $k$ can be normalized to unity. For $k>0,$ $k=0$
and $k<0$ we obtain respectively de Sitter, flat and anti-de Sitter spaces.
Thus, we just  demonstrated that the homogeneity and isotropy of space-time is
synonymous with spaces being de Sitter, flat and anti-de Sitter very much like
in Riemannian geometry there are spaces of positive, negative and zero
curvature. This observation can be used for obtaining  simple description of
\ just obtained results.

We begin by noticing that the surface of \ constant positive curvature is
conformally equivalent to a sphere embedded into flat Euclidean space [78,85].
In particular, let us consider a 3-sphere embedded into 4d Euclidean space. It
is described by the equation
\begin{equation}
S^{3}=\{x\in E_{4},\text{ }x_{1}^{2}+x_{2}^{2}+x_{3}^{2}+x_{4}^{2}%
=R^{2}\}.\tag{5.7}%
\end{equation}
$S^{3}$ is homogenous isotropic space with \ positive scalar curvature whose
value is $6/R^{2}.$ The group of motions associated with this homogenous space
is the rotation group $SO(4)$. The space of constant negative curvature
$H^{3}$ is obtained analogously. For this purpose it is sufficient, following
Dirac [78], to make $x_{1}$ purely imaginary and to replace $R^{2}$ by
$-R^{2}$ in (5.7). Such replacements produce:
\begin{equation}
H^{3}=\{x\in M_{4},\text{ }x_{1}^{2}-x_{2}^{2}-x_{3}^{2}-x_{4}^{2}%
=R^{2}\}.\tag{5.8}%
\end{equation}
In writing this result we have replaced the Euclidean space $E_{4}$ by
Minkowski space $M_{4}$ so that the rotation group $SO(4)$ is now replaced by
the Lorentz group $SO(3,1)$. The de Sitter space can now be obtained according
to Dirac as follows. In (5.7) we replace $E_{4}$ by $E_{5}$ and make $x_{1}$
purely imaginary \ thus converting $E_{5}$ into $M_{5}$. The obtained space is
the de Sitter space whose group of symmetry is $SO(4,1)$%
\begin{equation}
dS_{4}=\{x\in M_{5},\text{ }x_{1}^{2}-x_{2}^{2}-x_{3}^{2}-x_{4}^{2}-x_{5}%
^{2}=R^{2}\}.\tag{5.9}%
\end{equation}
It has a constant positive scalar curvature whose value is $12/R^{2}.$ Very
nice description of such a space is contained in the book by Hawking and Ellis
[88]. The connection between parameter $R$ and the cosmological constant
$\Lambda$ is given by $R=\sqrt{\dfrac{3}{\Lambda}}$. The anti-de Sitter space
of constant negative curvature is determined analogously. Specifically, it is
given by
\begin{equation}
adS_{4}=\{x\in E_{3,2},\text{ }x_{1}^{2}-x_{2}^{2}-x_{3}^{2}-x_{4}^{2}%
+x_{5}^{2}=R^{2}\},\tag{5.10}%
\end{equation}
where the five dimensional space $E_{3,2}$ is constructed by adding the
time-like direction to $M_{4}.$ Hence, the symmetry group of $adS_{4}$ is
SO(3,2). All these groups can be described simultaneously if, following Dirac
[78], we introduce the quadratic form
\begin{equation}%
%TCIMACRO{\tsum \limits_{\mu=1}^{5}}%
%BeginExpansion
{\textstyle\sum\limits_{\mu=1}^{5}}
%EndExpansion
x_{\mu}x_{\mu}=R^{2}\tag{5.11}%
\end{equation}
in which some of the arguments are allowed to be purely imaginary.
Transformations preserving such a quadratic form are appropriate respectively
for groups SO(5), SO(4,1) and SO(3,2). \ We can embed all these groups into
still a larger (conformal) group SO(4,2) by increasing summation from 5 to 6
in (5.11). In such a case all groups discussed in this work, starting from
SO(2,1), can be embedded into this conformal group as subgroups as discussed
in great detail by Wybourne [42]. Incidentally, the work by Graner and
Dubrulle entitled "Titius-Bode laws in the solar system I. Scale invariance
explains everything" [89],  when interpreted group-theoretically, becomes
\ just \ a corollary of such an embedding. The Titius-Bode law \ which these
authors reproduce by requiring the underlying system of equations to be
conformally invariant contains no restrictions \ on number of allowed orbit
discussed in Subection 4.1.2 \ Furthermore, their work requires many ad hoc
fitting assumptions. \ When using equations of fluid dynamics in their
subsequent work [90] to model evolution of the protoplanetary cloud of dust,
the obtained results contain only orbits of regular planets/satellites and,
hence suffers from the same type of problems as mentioned in the Introduction.
Uses of conformal symmetry in both gravity and conformal field theories has
been recently extended in our works [85,86]. The task still remains to find
out if representations of these larger groups \ can \ produce exact solutions
of \ the radial Schr\"{o}dinger equations not listed in the Natanzon-style
classification given in Ref.[44] for SO(2,1). If such solutions do exist, one
might be able to find those of them which are of relevance to celestial
quantum mechanics and, hence, to cosmology.

\section{Concluding remarks}

\ Although Einstein was not happy with the existing formulation of quantum
mechanics, the results presented in this work demonstrate harmonious
coexistence of general relativity and quantum mechanics to the extent that
existence of one implies existence of the other at the scales of our Solar
System.\ It should be noted though that such harmony had been achieved at the
expense of partial sacrificing of the correspondence principle. This principle
is not fully working anyway, even for such well studied system as Hydrogen
atom as discussed in Subsection 3.3.2. This fact is not too worrisome to us as
it was to Einstein. Indeed, as Heisenberg correctly pointed out: all what we
know about microscopic system is its spectrum (in the very best of cases). The
results of our recent works [15,16] as well as by mathematicians Knutson and
Tao [12-14] indicate that there are numerous ways to develop quantum
mechanics-all based on systematically analyzing combinatorics of the observed
spectral data. Such an approach is not intrinsic to quantum mechanics. In
works by Knutson and Tao  quantum mechanics was not discussed at all! Quantum
mechanical significance of their work(s) is discussed in detail in our recent
papers [15] While in [16] \ we developed quantum mechanical formalism based on
the theory of Poisson-Dirichlet-type processes. These stochastic processes are
not necessarily microscopic. Mathematically rigorous detailed exposition of
these processes is given in [91]. \ The combinatorial formalism developed in
[16] works equally well for quantum field and string theories. Not
surprisingly such formalism can be successfully applied to objects as big as
involved in\ the Solar System dynamics. Much more surprising is the unifying
role of gravity at the \textsl{microscopic} scales as discussed in our
\ latest work on gravity assisted solution of the mass gap
problem\footnote{This is one of the Millennium \ prize problems proposed by
the Clay Mathematics Institute, e.g. see
http://www.claymath.org/millennium/Yang-Mills\_Theory/} for Yang-Mills fields
[92].  Role of gravity in solution of the mass gap problem for Yang-Mills
fields is striking and unexpected. Although Einstein did not like quantum
mechanics because, as he believed, it is incompatible with his general
relativity, results of this work and those of reference [92] underscore the
profoundly deep connections between gravity and quantum mechanics/quantum
field theory at all scales. It is being hoped that our work will stimulate
\ development of more \ detailed expositions in the future, especially those
involving detailed study of \ spin-orbital interactions.

\bigskip\bigskip

\textbf{Appendix A. Some quantum mechanical problems associated with the Lie
algebra of} \textbf{SO(2,1)} \textbf{group}\bigskip

Following Wybourne [42] consider the second order differential equation of the
type%
\begin{equation}
\frac{d^{2}Y}{dx^{2}}+V(x)Y(x)=0\tag{A.1}%
\end{equation}
where $V(x)=a/x^{2}+bx^{2}+c.$ Consider as well the Lie algebra of the
noncompact group SO(2,1) or, better, its connected component SU(1,1). It is
given by the following commutation relations%
\begin{equation}
\lbrack X_{1},X_{2}]=-iX_{3};\text{ }[X_{2},X_{3}]=iX_{1};\text{ }[X_{3}%
,X_{1}]=iX_{2}\tag{A.2}%
\end{equation}
We shall seek the realization of this Lie algebra in terms of the following
generators%
\begin{equation}
X_{1}:=\frac{d^{2}}{dx^{2}}+a_{1}(x);\text{ \ }X_{2}:=i[k(x)\frac{d}{dx}%
+a_{2}(x)];\text{ \ }X_{3}:=\frac{d^{2}}{dx^{2}}+a_{3}(x).\tag{A.3}%
\end{equation}
The unknown functions $a_{1}(x),a_{2}(x),a_{3}(x)$ and $k(x)$ are determined
upon substitution of (A.3) into (A.2). After some calculations, the following
result is obtained%
\begin{equation}
X_{1}:=\frac{d^{2}}{dx^{2}}+\frac{a}{x^{2}}+\frac{x^{2}}{16};\text{ }%
X_{2}:=\frac{-i}{2}[x\frac{d}{dx}+\frac{1}{2}];\text{ }X_{3}:=\frac{d^{2}%
}{dx^{2}}+\frac{a}{x^{2}}-\frac{x^{2}}{16}.\tag{A.4}%
\end{equation}
In view of this, (A.1) can be rewritten as follows
\begin{equation}
\lbrack(\frac{1}{2}+8b)X_{1}+(\frac{1}{2}-8b)X_{3}+c]Y(x)=0.\tag{A.5}%
\end{equation}
This expression can be further simplified by the unitary transformation$UX_{1}%
U^{-1}=X_{1}\cosh\theta+X_{3}\sinh\theta;$ $UX_{3}U^{-1}=X_{1}\sinh
\theta+X_{3}\cosh\theta$ with $U=exp(-i\theta X_{2}).$ By choosing
$\tanh\theta=-(1/2+8b)/(1/2-8b)$ (A.5) is reduced to
\begin{equation}
X_{3}\tilde{Y}(x)=\frac{c}{4\sqrt{-b}}\tilde{Y}(x),\tag{A.6}%
\end{equation}
where the eigenfunction $\tilde{Y}(x)=UY(x)$ is an eigenfunction of both
$X_{3}$ and the Casimir operator \textbf{X}$^{2}=X_{3}^{2}-X_{2}^{2}-X_{1}%
^{2}$ so that by analogy with the Lie algebra of the angular momentum we
obtain,
\begin{align}
\mathbf{X}^{2}\tilde{Y}_{jn}(x) &  =J(J+1)\tilde{Y}_{Jn}(x)\text{
\ \ and}\tag{A.7a}\\
X_{3}\tilde{Y}_{Jn}(x) &  =\frac{c}{4\sqrt{-b}}\tilde{Y}_{Jn}(x)\equiv
(-J+n)\tilde{Y}_{Jn}(x)\text{; }\ n=0,1,2,...\text{.}\tag{A.7b}%
\end{align}
It can be shown that $J(J+1)=-a/4-3/16$. From here we obtain : $J=-\frac{1}%
{2}(1\pm\sqrt{\frac{1}{4}-a});$ $\frac{1}{4}-a\geq0.$ In the case of discrete
spectrum one should choose the plus sign in the expression for $J$. Using this
result in (A.7) we obtain the following result of major importance%

\begin{equation}
4n+2+\sqrt{1-4a}=\frac{c}{\sqrt{-b}}.\tag{A.8}%
\end{equation}
Indeed, consider the planar Kepler problem. In this case, in view of (3.5),
the radial Schr\"{o}dinger equation can be written in the following symbolic
form%
\begin{equation}
\left[  \frac{d^{2}}{dr^{2}}+\frac{1}{r}\frac{d}{dr}+\frac{\mathit{\upsilon}%
}{r}+\frac{u}{r^{2}}+g\right]  R(r)=0\tag{A.9}%
\end{equation}
By writing $r=x^{2}$ and $R(r)=x^{-\frac{1}{2}}\mathcal{R}(x)$ This equation
is reduced to the canonical form given by (A.1), e.g. to%
\begin{equation}
(\frac{d^{2}}{dx^{2}}+\frac{4u+1/4}{x^{2}}+4gx^{2}+4\upsilon)\mathcal{R}%
(x)=0\tag{A.10}%
\end{equation}
so that the rest of arguments go through. Analogously, in the case of
Morse-type potential we have the following Schrodinger-type equation
initially:%
\begin{equation}
\left[  \frac{d^{2}}{dz^{2}}+pe^{2\alpha z}+qe^{\alpha z}+k\right]
R(z)=0\tag{A.11}%
\end{equation}
By choosing $z=lnx^{2}$ and $R(z)=x^{-\frac{1}{2}}\mathcal{R}(x)$ (A11) is
reduced to the canonical form%
\begin{equation}
(\frac{d^{2}}{dx^{2}}+\frac{16k+\alpha^{2}}{4\alpha^{2}x^{2}}+\frac{4p}%
{\alpha^{2}}x^{2}+\frac{4q}{\alpha^{2}})\mathcal{R}(x)=0.\tag{A.12}%
\end{equation}
By analogous manipulations one can reduce to the canonical form the radial
equation for  Hydrogen atom and for 3-dimensional harmonic oscillator.

\bigskip

\textbf{Appendix B}. \textbf{Numerical data used for claculations of}
\textbf{n}$_{theory}^{\ast}$

( \textbf{Supplement to} \textbf{Table 2}).

\ 

1 au=149.598$\cdot10^{6}km$

\smallskip

Masses (in kg): Sun 1.988$\cdot10^{30},$ Jupiter 1.8986$\cdot10^{27}$, Saturn
5.6846$\cdot10^{26},$

Uranus 8.6832$\cdot10^{25},$ Neptune 10.243$\cdot10^{25}.$

\smallskip

q$_{j}:$ Jupiter 0.955$\cdot10^{-3},$ Saturn 2.86$\cdot10^{-4},$ Uranus
4.37$\cdot10^{-5}$, Neptune 5.15$\cdot10^{-5}.$

\smallskip

$\left(  r_{j}\right)  _{1}(km):$ Jupiter 127.69$\cdot10^{3},$ Saturn
133.58$\cdot10^{3},$ Uranus 49.77$\cdot10^{3},$

Neptune 48.23$\cdot10^{3}.$

\smallskip

ln$\left(  \dfrac{\gamma M}{2r_{1}}\right)  $ : Earth 4.0062, Jupiter 3.095,
Saturn 1.844, Uranus 0.9513,

Neptune 1.15.

\medskip

\ \textbf{Appendix C}. \textbf{Connections between the Hill and Neumann's }

\textbf{dynamical problems.}

\ 

We follow our paper [93] where some mathematical of the results of the paper
by Lazutkin and Pankratova (1975) were used\ for solution of concrete physical
problems. In particular, following our paper, let us consider the
Fuchsian-type equation given by
\begin{equation}
y^{^{\prime\prime}}+\frac{1}{2}\phi y=0,\tag{C.1}%
\end{equation}
where the potential $\phi$ is determined by the equation $\phi=[f]$ with
$f=y_{1}/y_{2}$ and $y_{1},y_{2}$ \ being two independent solutions of (C.1)
normalized by the requirement $y_{1}^{^{\prime}}y_{2}$ -$y_{2}^{\prime}%
$\ $y_{1}=1.$The symbol$\ [f]$ denotes the Schwarzian derivative of $f$. Such
a derivative is defined as follows
\begin{equation}
\lbrack f]=\frac{f^{\prime}f^{\prime\prime\prime}-\frac{3}{2}\left(
f^{\prime\prime}\right)  ^{2}}{\left(  f^{\prime}\right)  ^{2}}.\tag{C.2}%
\end{equation}
Consider (C.1) on the circle $S^{1}$ and consider some map of the circle given
by $F(t+1)=F(t)+1.$ Let $t=F(\xi)$ so that $y(t)=Y(\xi)\sqrt{F^{\prime}(\xi)}$
leaves (C.1) form -invariant, i.e. in the form $Y^{\prime\prime}+\frac{1}%
{2}\Phi Y=0$ with potential $\Phi$ being defined now as $\Phi(\xi)=\phi
(F(\xi))[F^{\prime}(\xi)]^{2}+[F(\xi)].$ Consider next the infinitesimal
transformation $F(\xi)=\xi+\delta\varphi(\xi)$ with $\delta$ being some small
parameter and $\varphi(\xi)$ being some function to be determined. Then,
$\Phi(\xi+\delta\varphi(\xi))=\phi(\xi)+\delta(\hat{T}\varphi)(\xi
)+O(\delta^{2}).$ Here $(\hat{T}\varphi)(\xi)=\phi(\xi)\varphi^{\prime}%
(\xi)+\frac{1}{2}\varphi^{\prime\prime\prime}(\xi)+2\phi^{\prime}(\xi
)\varphi(\xi).$ Next, we assume that the parameter $\delta$ plays the same
role as time. Then, we obtain
\begin{equation}
\lim_{t\rightarrow0}\frac{\Phi-\phi}{t}=\frac{\partial\phi}{\partial t}%
=\frac{1}{2}\varphi^{\prime\prime\prime}(\xi)+\phi(\xi)\varphi^{\prime}%
(\xi)+2\phi^{\prime}(\xi)\varphi(\xi)\tag{C.3}%
\end{equation}
Since thus far the perturbing function $\varphi(\xi)$ was left undetermined,
we can choose it now as $\varphi(\xi)=\phi(\xi).$ Then, we obtain the Korteweg
-de Vriez \ (KdV) equation%
\begin{equation}
\frac{\partial\phi}{\partial t}=\frac{1}{2}\phi^{\prime\prime\prime}%
(\xi)+3\phi(\xi)\phi^{\prime}(\xi)\tag{C.4}%
\end{equation}
determining the potential $\phi(\xi).$ For reasons which are explained in the
text, it is sufficient to consider only the static case of KdV, i.e.%
\begin{equation}
\phi^{\prime\prime\prime}(\xi)+6\phi(\xi)\phi^{\prime}(\xi)=0.\tag{C.5}%
\end{equation}
We shall use this result as a reference for our main task of connecting the
Hill and the Neumann's problems. Using (4.19) we write%
\begin{equation}
u(\xi)=\phi(\xi)-<\dot{\xi},\dot{\xi}>.\tag{C.6}%
\end{equation}
Consider an auxiliary functional $\varphi(\xi)=<\xi,A^{-1}\xi>.$ Suppose that
$\varphi(\xi)=u(\xi).$ Then,
\begin{equation}
\frac{du}{dt}=2<\dot{\xi},A\xi>-2<\ddot{\xi},\dot{\xi}>.\tag{C.7}%
\end{equation}
But $<\ddot{\xi},\dot{\xi}>=0$ because of the normalization constraint
$<\xi,\xi>=1.$ Hence, $\dfrac{du}{dt}=2<\dot{\xi},A\xi>.$ Consider as well
$\dfrac{d\varphi}{dt}.$ By using (4.19) it is straightforward to show that
$\dfrac{d\varphi}{dt}=2<\dot{\xi},A^{-1}\xi>.$ Because by assumption
$\varphi(\xi)=u(\xi),$ we have to demand that $<\dot{\xi},A^{-1}\xi>=<\dot
{\xi},A\xi>$ as well. If this is the case, consider
\begin{equation}
\dfrac{d^{2}u}{dt^{2}}=2<\ddot{\xi},A^{-1}\xi>+2<\dot{\xi},A^{-1}\dot{\xi
}>.\tag{C.8}%
\end{equation}
Using (4.19) once again we obtain,%
\begin{equation}
\dfrac{d^{2}u}{dt^{2}}=-2+2u\varphi+2<\dot{\xi},A^{-1}\dot{\xi}>.\tag{C.9}%
\end{equation}
Finally, consider as well $\dfrac{d^{3}u}{dt^{3}}.$ Using (C.9) as well as
(4.19) and (C.7) we obtain,%
\begin{equation}
\dfrac{d^{3}u}{dt^{3}}=2\frac{du}{dt}\varphi+4u\frac{du}{dt}=6u\frac{du}%
{dt}.\tag{C.10}%
\end{equation}
By noticing that in (C.5) we can always make a rescaling $\phi(\xi
)\rightarrow\lambda\phi(\xi),$ we can always choose $\lambda=-1$ so that (C.5)
and (C.10) coincide. This result establishes correspondence between the
Neumann and Hill-type problems.

QED

\bigskip\bigskip

\textbf{References}

\bigskip\bigskip

\ [1] \ \ \vspace{0in}Manin Y 2010 \ AMS Notices \textbf{57} 239

\ [2] \ \ Kac V and Cheung P 2002 \textsl{Quantum Calculus}

\ \ \ \ \ \ \ (Berlin: Springer-Verlag)

\ [3] \ \ Porter M and Cvitanovi\v{c} P 2005 AMS Notices \ \textbf{52 }1020-1025

\ [4] \ \ Marsden J and Ross S 2006 AMS Bulletin \textbf{43} 43-73

\ [5] \ \ Convay B, Chilan C and Wall B 2007

\ \ \ \ \ \ \ Cel. Mech. Dynam. Astron. \ \textbf{97}, 73-86

\ [6] \ \ Poincare$^{\prime}$ H 1892-1898 \textsl{Les Methodes Nouvelles de
la}

\ \ \ \ \ \ \ \textsl{Mechanique Celeste} (Paris: Gauthier-Villars)

\ [7] \ \ Woollfson M 2007 The Formation of the Solar System

\ \ \ \ \ \ \ (Singapore: World Scientific)

\ [8] \ \ Tiscareno M and Hedman M 2009 Nature \textbf{461}1064

\ [9] \ \ Verbiscer A, Skrutskie M and Hamilton D 2009 Nature \textbf{461}1098

[10] \ \ Celletti A and Perozzi E 2007 \textsl{Celestial Mechanics}.

\ \ \ \ \ \ \ \textsl{The Waltz of the Planets} (Berlin: Springer)

[11] \ \ Murray C and Dermott S 1999 \textsl{Solar System Dynamics}.

\ \ \ \ \ \ \ (Cambridge UK: Cambridge University Press)

[12] \ \ Knutson A and Tao\ T 2001 AMS Notices \ \textbf{48 }175

[13] \ \ Knutson A and Tao T 1999 \ AMS Journal \textbf{12}1055

[14] \ \ Knutson A, Tao T and Woodward,C 2004

\ \ \ \ \ \ \ \ AMS Journal \textbf{17} 19

[15] \ \ Kholodenko A 2009 Math.Forum 2009 \textbf{4} 441

[16] \ \ Kholodenko A 2008 EJTP \textbf{5} 35 \ arXiv:0806.1064

[17] \ \ Einstein A 1966 Collected Works Vol.4 (Moscow: Nauka)

\ \ \ \ \ \ \ \ in Russian

[18] \ \ Sanders R 2010 UC Berkeley News 02 17

[19] \ \ Reynaud S and Jaekel M 2008 arXiv:0801.3407

[20] \ \ Goldstein H, Poole C and Safko J 2002 \textsl{Classical Mechanics}

\ \ \ \ \ \ \ \ (New York: Addison-Wesley)

[21] \ \ Landau L and Lifshitz E. 1975 \textsl{The Classical Theory of Fields}

\ \ \ \ \ \ \ \ (London: Pergamon)

[22] \ \ Infeld L and Schield A 1949 Rev.Mod.Phys. \textbf{21} 408

[23] \ \ Pound A 2010 Phys.Rev.D \textbf{81} 024023

[24] \ \ Laplace P 1966 Celestial Mechanics Vols 1-4

\ \ \ \ \ \ \ \ (Bronx, NY: Chelsea Pub. Co.)

[25] \ \ Arnol'd V, KozlovV. and Neishtadt A 2006 \textsl{Mathematical
Aspects}

\ \ \ \ \ \ \ \ \textsl{of Classical and Celestial Mechanics}

\ \ \ \ \ \ \ \ (Heidelberg: Springer)

[26] \ \ Heisenberg W 1925 Z.Phys. \textbf{33}, 879 (in German)

[27] \ \ Charlier C 1927 \textsl{Die Mechanik des Himmels}

\ \ \ \ \ \ \ \ (Berlin:Walter de Gruyter) in German

[28]\ \ \ Fejoz J 2004 Erg.Th.\&Dyn.Syst. \textbf{24} 1521

[29] \ \ Biasco L, Cherchia L, Valdinoci E 2006

\ \ \ \ \ \ \ \ SIAM J.Math.Anal. \textbf{37} 1580

[30] \ \ \ Moser J and Zehnder E 2005 \textsl{Notes on Dynamical Systems}.

\ \ \ \ \ \ \ \ (Providence, RI :\ AMS Publishers)

[31] \ \ \ Stephani H 1990 General Relativity

\ \ \ \ \ \ \ \ (Cambridge UK: Cambridge University Press)

[32] \ \ \ Doi M and Edwards S 1986 \textsl{The Theory of Polymer Dynamics}

\ \ \ \ \ \ \ \ (Oxford: Clarendon)

[33] \ \ \ Kholodenko A, Ballauff M and Granados M 1998 Physica \textbf{A 260} 267

[34] \ \ \ Dirac P 1926 Proc.Roy.Soc. A \textbf{109} 642

[35] \ \ \ Dirac P 1958 \textsl{Principles of Quantum Mechanics}

\ \ \ \ \ \ \ \ (Oxford: Clarendon Press,)

[36] \ \ \ Pars L 1968 \textsl{Analytical Dynamics} (London: Heinemann)

[37] \ \ \ Dirac P 1950 Canadian J.Math. \textbf{2}, 129

[38] \ \ \ Bander M and Itzykson C 1966 \ Rev. Mod. Phys. \textbf{38} 330

[39] \ \ \ Kac M 1966 \ Am.Math.Monthly \textbf{73} 1

[40] \ \ \ Dhar A, Rao D, Shankar U and Sridhar S 2003

\ \ \ \ \ \ \ \ \ Phys.Rev.E \textbf{68} 026208

[41] \ \ \ Jauch J and Hill E 1940 Phys.Rev. \textbf{57} 641

[42] \ \ \ Wybourne B 1974 \textsl{Classical Groups for Physicists}

\ \ \ \ \ \ \ \ \ (New York: John Willey \& Sons)

[43] \ \ \ Natanzon G 1979 Theor.Math.Phys. \textbf{38} 219

[44] \ \ \ Levai G 1994 J.Phys.A \textbf{27}, 3809

[45] \ \ \ Cordero P, Holman S, Furlan P and Ghirardi G 1971

\ \ \ \ \ \ \ \ \ Il Nuovo Cim. \textbf{3A} 807

[46] \ \ \ Bargmann V 1947 Ann.Math. \textbf{48} 568-640 (1947)

[47] \ \ \ Barut A and Fronsdal C 1965 Proc.Roy.Soc.London \textbf{A}
\textbf{287} 532

[48] \ \ \ Cooper F, Ginoccio J and Khare A 1987 Phys.Rev.\textbf{D 36} 2458

[49] \ \ \ Junker G and Roy P 1998 Ann.Phys. \textbf{270} 155

[50]\ \ \ \ Boyer T 1975 Phys.Rev.D \textbf{11} 790

[51] \ \ \ Puthoff H 1987 Phys.Rev. \textbf{D 35} 3266

[52] \ \ \ Goldreich P 1965 Mon. Not.R. Astr.Soc. \textbf{130} 159

[53] \ \ \ Ortin T 2004 \textsl{Gravity and Strings}

\ \ \ \ \ \ \ \ (Cambridge UK: Cambridge University Press)

[54] \ \ \ Nieto M 1972 \textsl{The Titius -Bode Law of Planetary Distances:}

\ \ \ \ \ \ \ \ \textsl{Its History and Theory} (London: Pergamon Press)

[55] \ \ \ Neslu\v{s}an L 2004 Mon. Not.R.Astr.Soc. \textbf{351} 133

[56] \ \ \ Lynch P 2003 Mon.Not. R.Astr.Soc. \textbf{341} 1174

[57] \ \ \ Dermott S 1968 Mon.Not.R.Astr.Soc. \textbf{141} 363

[58] \ \ \ Hayes W andTremaine S 1998 Icarus \textbf{135} 549

[59] \ \ \ Landau L and Lifshitz E 1962 \textsl{Quantum Mechanics}

\ \ \ \ \ \ \ \ (London:Pergamon)

[60] \ \ \ Khriplovich I 2005 \textsl{General Relativity}

\ \ \ \ \ \ \ \ \ (Berlin:Springer)

[61] \ \ \ Papapetrou A 1951 Proc.Roy.Soc.A \textbf{209} 248

[62] \ \ \ Dermott S 1968 Mon.Not.R.Astr.Soc. \textbf{141} 363

[63] \ \ \ Chebotarev G 1968 \textsl{Analytical and Numerical Methods of
Celestial}

\ \ \ \ \ \ \ \ \textsl{Mechanics} (Amsterdam: Elsevier)

[64] \ \ \ SzebehelyV 1967 \textsl{Theory of Orbits}

\ \ \ \ \ \ \ \ (New York: Academic Press)

[65] \ \ \ Avron J and Simon B 1981 PRL \textbf{46} 1166

[66] \ \ \ Ince E 1926 \textsl{Ordinary Differential Equations}

\ \ \ \ \ \ \ \ (New York: Dover Publishers)

[67] \ \ \ MagnusW and Winkler S 1966 \textsl{Hill's Equation}

\ \ \ \ \ \ \ \ (New York : Interscience Publishers)

[68] \ \ \ Mc Kean, H and\ Moerbeke P 1975 Inv.Math. \textbf{30} 217

[69] \ \ \ Trubowitz, E 1977 Comm. Pure Appl. Math. \textbf{30} 321

[70]\ \ \ \ Moser J 1980 \ In: \textsl{Dynamical Systems}, pp.233-290

\ \ \ \ \ \ \ \ \ (Boston: Birkh\"{a}user)

[71] \ \ \ Hochstadt H 1963 Math.Zeitt. \textbf{82} 237

[72] \ \ \ Esposito L 2006 \textsl{Planetary Rings} Cambridge University Press.

\ \ \ \ \ \ \ \ \ (Cambridge, UK : Cambridge University Press)

[73] \ \ \ Kirillov A 1982 \textsl{Infinite dimensional Lie groups: their
orbits},

\ \ \ \ \ \ \ \ \textsl{\ invariants and \ representations}
LNM\ \ \textbf{970}, 101

[74] \ \ \ Arnol'd V and Khesin B 1998 \textsl{Topological Methods in }

\ \ \ \ \ \ \ \ \textsl{Hydrodynamics }\ (Berlin: Springer)

[75] \ \ \ Vilenkin N 1991 \textsl{Special Functions and Theory of Group}

\ \ \ \ \ \ \ \ \textsl{\ Representations }(Moscow: Nauka) in Russian

[76] \ \ \ \ Veselov A, Dullin H, Richter P and Waalkens H 2001

\ \ \ \ \ \ \ \ \ Physica \textbf{D} \textbf{155} 159

[77] \ \ \ Einstein A, Infeld L and Hoffmann B 1938

\ \ \ \ \ \ \ \ \ Ann.Math. \textbf{39} 65

[78] \ \ \ Dirac P 1935 Ann.Math. \textbf{36} 657

[79] \ \ \ Polyakov A 2009 arXiv:0912.5503

[80] \ \ \ Bros J, Epstein H and Moschella U 2006

\ \ \ \ \ \ \ \ \ arxiv: hep-th/0612184

[81] \ \ \ Peebles P and Ratra B 2003

\ \ \ \ \ \ \ \ \ Rev.Mod Phys.\textbf{75} 559

[82] \ \ \ Copeland E, Sami M and Tsujikawa S 2006

\ \ \ \ \ \ \ \ \ arxiv: hep-th/0603057

[83] \ \ \ Kay S, Pearce F, Frenk C and Jenkins A 2002

\ \ \ \ \ \ \ \ \ Mon.Not.R. Astron. Soc. \textbf{330} 113

[84] \ \ \ De Blok W, McGauh S, Bosma A and RubinV 2001

\ \ \ \ \ \ \ \ \ Astrophys. J.Lett. \textbf{552} L 23

[85] \ \ \ Kholodenko A 2008 J.Geom.Phys. \textbf{58} 259

[86] \ \ \ Kholodenko A and Ballard E 2007 Physica A \textbf{380} 115

[87] \ \ \ Willmore T 1993 \textsl{Riemannian Geometry}

\ \ \ \ \ \ \ \ \ (Oxford :Clarendon Press)

[88] \ \ \ Hawking S and Ellis G 1973\textsl{\ The Large Scale Structure}

\ \ \ \ \ \ \ \ \ \textsl{of Space-Time }(Cambridge UK:Cambridge University Press)

[89] \ \ \ Graner F and Dubrulle B 1994 Astron.Astrophys. \textbf{282} 262

[90] \ \ \ Graner F and Dubrulle B 1994 Astron.Astrophys. \textbf{282} 269

[91] \ \ \ Bertoin J 2006 \textsl{Random Fragmentation and Coagulation}

\ \ \ \ \ \ \ \ \ \textsl{Processes }(Cambridge, UK : Cambridge University Press)

[92] \ \ \ \ Kholodenko A 2010 arXiv:1001.0029

[93] \ \ \ \ Kholodenko A 2002 \ J.Geom.Phys. \textbf{43} 45

\ \ \ \ \ \ \ \ 

\ \ \ \ \ \ 

\bigskip

\ \ \ \ \ \ \ \ 

\ \ \ \ \ \ \ 

\ \ \ \ \ \ \ 

\ \ \ \ 

\bigskip

\bigskip
\end{document}